\documentclass[aps,graphicx,prb,twocolumn,groupedaddress,showpacs,notitlepage]{revtex4-1}  
\usepackage{graphicx}
\usepackage{amsmath,amssymb}
\usepackage{color}

\newcommand{\bra}[1]{\left<#1\right|}
\newcommand{\ket}[1]{\left|#1\right>}
\newcommand{\sh}[0]{{\rm sh}\;}
\newcommand{\ch}[0]{{\rm ch}\;}
\newcommand{\abs}[1]{\left|#1\right|}
\newcommand{\Tr}[1]{\text{Tr}\left(#1\right)}
\newcommand{\av}[1]{\left<#1\right>}

\begin{document}
\title{Theory of Raman and Resonant Inelastic X-ray Scattering from
  Collective Orbital Excitations in YTiO$_3$}
\author{L. J. P. Ament}
\affiliation{Institute-Lorentz for Theoretical Physics, Universiteit Leiden, P.O. Box 9506, 2300 RA Leiden, The Netherlands}
\author{G. Khaliullin}
\affiliation{Max-Planck-Institut f\"ur Festk\"orperforschung, Heisenbergstrasse 1, D-70569 Stuttgart, Germany}
\date{\today}

\begin{abstract}
We present two different theories for Raman scattering and Resonant Inelastic X-ray
Scattering (RIXS) in the low temperature ferromagnetic phase of YTiO$_3$ and
compare this to the available experimental data. 
For description of the orbital ground-state and orbital excitations, 
we consider two
models corresponding to two theoretical limits: one where the $t_{2g}$
orbitals are degenerate, and the other where strong lattice distortions split
them. In the former model the orbitals interact through superexchange. The
resulting superexchange Hamiltonian yields an orbitally ordered ground state
with collective orbital excitations on top of it -- the orbitons. In the
orbital-lattice model, on the other hand, distortions lead to local
$dd$-transitions between crystal field levels. Correspondingly, the 
orbital response functions that determine Raman and RIXS lineshapes and
intensities are of cooperative or single-ion character.  
We find that the superexchange model yields theoretical Raman and RIXS 
spectra that fit very well to the experimental data. 
\end{abstract}


\maketitle

\begin{section}{Introduction}
In many 3$d$ transition metal compounds, like the titanates and colossal magnetoresistance manganites, the low-energy orbital degrees of freedom are (approximately) degenerate. This leads to all kinds of interesting phenomena like cooperative Jahn-Teller (JT) distortions, orbital frustration and strong spin-orbit coupling. Orbitals on neighboring orbitally active ions can be coupled via JT-distortions and via superexchange\cite{Gehring1975,Kugel1982,Goodenough1963}. Both couplings can give rise to an orbitally ordered ground state. However, the nature of the orbital excitations on top of the orbitally ordered ground state is very different depending on the coupling mechanism. In a pure superexchange system, collective excitations called `orbitons' emerge. These collective orbital excitations are coherent waves of excited orbitals, similar to the spin wave excitations in spin systems. When this system is coupled to the lattice, as is the case when JT-distortions are present, interaction with phonons destroys the orbiton coherence. The distortions lift the orbital degeneracy and the excitations become local $dd$-excitations.

The titanates, with a pseudo-cubic perovskite lattice structure, are
good candidates to support orbitons. The Ti ions with their 3$d^1$
configuration have one electron in one of the three nearly degenerate
$t_{2g}$ orbitals. Since these orbitals are directed away from the
neighboring oxygen ions, the coupling to the lattice is expected to be
small. Further, it has been shown\cite{Khaliullin2002,Khaliullin2003}
that a superexchange-only model explains many of the ground state
properties of YTiO$_3$. Also, there is experimental evidence that
LaTiO$_3$ is a (superexchange-driven) orbital
liquid\cite{Keimer2000,Khaliullin2000,Cheng2008}. On the other hand,
local crystal field models also well reproduce some of the physical
properties of the titanates\cite{Mochizuki2003,Pavarini2004,Pavarini2005,
Cwik2003,Schmitz2005,Haverkort2005,Solovyev2006,Akimitsu2001,Kiyama2003,
Kubota2004,Kiyama2005}. Both
models have their shortcomings as well: a Jahn-Teller dominated
description is not able to reproduce the spin wave spectrum, which is
nearly isotropic in both spin and real space, while the superexchange
model has difficulties explaining the experimentally observed orbital
polarization\cite{Akimitsu2001,Kiyama2003,Kubota2004,Kiyama2005}. Consequently,
it still remains controversial which mechanism dominates the orbitals
in titanates\cite{Khaliullin2005}.

In order to resolve this controversy, it is of crucial importance to compare recent Raman and Resonant Inelastic X-ray Scattering (RIXS) experiments on titanates\cite{Ulrich2006,Ulrich2008,experimentalpaper} to both of the competing theories.

The experimental observation of orbitons is a difficult task, as is underscored by the heavily debated Raman measurements on LaMnO$_3$\cite{Saitoh2001,Grueninger2002,Saitoh2002}. Because the hallmark of collective excitations is dispersion, a much better technique to directly probe orbitons is RIXS. With its energy resolution drastically improved over the last few years, RIXS now offers a whole new way of accessing the elementary excitations of solids, complementary to for instance Raman and neutron scattering. In this paper we analyze recent Raman and RIXS spectra\cite{Ulrich2006,Ulrich2008,experimentalpaper} for YTiO$_3$ from the point of view of a superexchange-only model and the alternative extreme of a completely local, lattice distortion dominated model. We find that while the orbital-lattice model can be finetuned to capture some aspects of the observed spectra, the collective superexchange model yields a much better overall description of the Raman and RIXS data.

The paper is organized as follows: Sec.~\ref{sec:titanates} compactly
reviews previous work on YTiO$_3$
and introduces the superexchange formalism and the
 local crystal field model. Sections~\ref{sec:raman} and
\ref{sec:rixs} deal with the theory of Raman scattering and RIXS
respectively, in both the superexchange and crystal field models.
\end{section}

\begin{section}{Two models of YTiO$_3$ \label{sec:titanates}}
For the existence of collective excitations of orbitals, the so-called orbitons, it makes a difference whether the orbital order is driven by JT distortions or superexchange\cite{Gehring1975,Kugel1982}. For large JT distortions, the crystal field splitting is large and a local picture applies: the collective nature of the orbital excitations characteristic of orbitons is lost. In materials where the orbital-lattice coupling is small, the superexchange interactions between orbitals can dominate over crystal field splittings due to lattice distortions. The Ti ions have a 3$d^1$ configuration, and the octahedral crystal field induces a splitting between the higher energy $e_g$ and lower energy $t_{2g}$ levels. Because the $t_{2g}$ orbitals are not directed towards neighboring oxygen ions, they are not expected to couple strongly to lattice distortions.

Building on this assumption, one can derive a superexchange Hamiltonian
starting from a Hubbard model. Below, we follow
Refs.~[\onlinecite{Khaliullin2002,Khaliullin2003}] closely. By symmetry, the
hopping term connects, for instance, the $zx$ to $zx$ and $yz$ to $yz$
orbitals along the $z$-direction ($c$-axis) via the intermediate 
oxygen 2$p_{\pi}$
states. $xy$ orbitals are not coupled along this direction. In the limit of
large on-site Coulomb repulsion $U$, this leads to a superexchange interaction
that depends on the spatial direction of a bond, and the resulting model is
intrinsically frustrated: on any given ion, there is no orbital that minimizes
the bond energy in all directions simultaneously. 

Because YTiO$_3$ is ferromagnetic at low temperature\cite{Goral1982} ($T_c \approx 30$ K), we restrict ourselves to the completely ferromagnetic part of the Hilbert space. Then one obtains the simple Hamiltonian
\begin{equation}
  \hat{H}_0 = \frac{1}{2} J_{\rm orb} \sum_{\av{i,j}} \left(
  \hat{A}^{(\gamma)}_{ij} +\frac{n_{\gamma,i}+n_{\gamma,j}}{2}
  \right), \label{eq:H}
\end{equation}
with the orbital exchange integral $J_{\rm orb} = r_1 J_{SE}$, where $r_1 = 1/(1-3 J_H/U) \approx 1.56$ parametrizing Hund's rule coupling and $J_{SE} = 4t^2/U$ is the superexchange constant derived from the Hubbard model. The operator $\hat{A}^{(\gamma)}_{ij}$ depends on the direction $\gamma$ of the bond $ij$. For example, in the $z$-direction we have
\begin{equation}
	\hat{A}^{(c)}_{ij} = n_{a,i}n_{a,j} + n_{b,i}n_{b,j} + a^{\dag}_i
        b^{\phantom{\dag}}_i b^{\dag}_j a^{\phantom{\dag}}_j + b^{\dag}_i
        a^{\phantom{\dag}}_i a^{\dag}_j b^{\phantom{\dag}}_j. 
\label{eq:A}
\end{equation}
The operators $a^{\dag}, b^{\dag}$ and $c^{\dag}$ create an electron
in the $yz$-, $zx$- and $xy$-orbital, respectively, and $n_a =
a^{\dag} a$. The Hamiltonian can also be written in terms of
interacting effective angular momenta $l=1$, operating on the $t_{2g}$
triplet. Because of the orbital frustration, these can form a myriad
of different classical ground
states. Refs.~[\onlinecite{Khaliullin2002}, \onlinecite{Khaliullin2003}] conclude that a $4$-sublattice quadrupole ordered state is favored, in which the orbitals
\begin{equation}
	\ket{\psi_c} = \frac{1}{\sqrt{3}} \left( \ket{d_{yz}} \pm \ket{d_{zx}}
          \pm \ket{d_{xy}} \right)
\label{eq:psiSE}
\end{equation}
are condensed. The signs $\pm$ alternate between the sublattices, such
that nearest-neighbor orbitals are orthogonal, supporting
ferromagnetic order. On top of this condensate, two species of
orbitons can be created, loosely speaking by populating either one of
the two orbitals orthogonal to $\psi_c$. The orbiton spectrum has
$3N^{1/3}$ Goldstone modes (where $N$ is the total number
of Ti ions), because the number of orbitals of a specific ``color'' is conserved in the plane in which it is lying. However, in YTiO$_3$ the TiO$_6$ octahedra are tilted. Because of this, hopping between different $t_{2g}$ orbitals is now no longer symmetry forbidden, and the conservation of orbital ``color'' is violated, removing the Goldstone modes. When also some anharmonic terms of the Hamiltonian are taken into account on a mean field level, the orbiton dispersion becomes\cite{Khaliullin2003}
\begin{align}
	\omega_{1/2,{\bf k}} = &\sqrt{Z_{\varepsilon} Z_f} J_{\rm orb} \{ 1-(1-2\varepsilon)(1-2f)(\gamma_{1,{\bf k}} \pm \kappa_{\bf k})^2 \nonumber \\
	&-2(\varepsilon -f)(\gamma_{1,{\bf k}} \pm \kappa_{\bf k} ) \}^{1/2}, \label{eq:dispersion}
\end{align}
where we use the signs $+$ and $-$ for $\omega_{1,{\bf k}}$ and $\omega_{2,{\bf k}}$ respectively. Further, $\sqrt{Z_{\varepsilon} Z_f} \approx 1.96$, $f \approx 0.086$, $\varepsilon \approx 0.18$, $\gamma_{1,{\bf k}} = (c_x+c_y+c_z)/3$ and $\kappa_{\bf k} = \sqrt{\gamma^2_{2,{\bf k}} + \gamma^2_{2,{\bf k}}}$ with $\gamma_{2,{\bf k}} = \sqrt{3}(c_y-c_x)/6$ and $\gamma_{3,{\bf k}} = (2 c_z-c_x-c_y)/6$ with $c_{\alpha} = \cos k_{\alpha}$. Eq.~(\ref{eq:dispersion}) describes the collective orbital modes that disperse up to energies of $2J_{\rm orb}$ and have a gap of approximately $J_{\rm orb}$.

In the second orbital model for YTiO$_3$ that we consider, lattice distortions dominate over superexchange interactions. Pavarini {\it et al.}\cite{Pavarini2004,Pavarini2005} did a DMFT+LDA calculation and found that lattice distortions of the GdFeO$_3$-type lift the orbital degeneracy. They also obtained four sublattices. The resulting local eigenstates of the $t_{2g}$ system are\cite{Pavarini2005}
\begin{align}
	\ket{1} &= 0.781 \ket{yz} -0.073 \ket{zx} + 0.620 \ket{xy} \label{eq:local1} \\
	\ket{2} &= -0.571 \ket{yz} +0.319 \ket{zx} + 0.757 \ket{xy} \\
	\ket{3} &= 0.253 \ket{yz} +0.945 \ket{zx} - 0.207 \ket{xy} \label{eq:local3}
\end{align}
for sublattice 1, with corresponding orbital energies $\epsilon_1 =
289$ meV, $\epsilon_2 = 488$ meV and $\epsilon_3 = 620$ meV. This
yields excitation energies $\omega_1 = \epsilon_2-\epsilon_1 = 199$
meV, $\omega_2 = \epsilon_3-\epsilon_1 = 331$ meV. The orbital states
on the other sublattices can be obtained from lattice symmetry
considerations\cite{Pavarini2005}. Superexchange processes are treated
as a perturbation in this model, broadening the states generated by
lattice distortions. This picture is also supported by other
theoretical work\cite{Mizokawa1996,Mochizuki2003,Schmitz2005,Ishihara2004}. 

It is possible to rotate the axes on each of the sublattices in such a way
that in the new coordinates, the eigenstates are still given by
Eqs.~(\ref{eq:local1}) through (\ref{eq:local3}): 
\begin{align}
	{\rm subl.}\; 1: (x,y,z) &\mapsto (x,y,z) \\
	{\rm subl.}\; 2: (x,y,z) &\mapsto (y,x,z) \\
	{\rm subl.}\; 3: (x,y,z) &\mapsto (x,y,-z) \\
	{\rm subl.}\; 4: (x,y,z) &\mapsto (y,x,-z).
\end{align}
Correspondingly, the orbiton operators transform as follows:
\begin{align}
	{\rm subl.}\; 1: (a,b,c) &\mapsto (a,b,c) \label{eq:localrotate1}\\
	{\rm subl.}\; 2: (a,b,c) &\mapsto (b,a,c) \\
	{\rm subl.}\; 3: (a,b,c) &\mapsto (-a,-b,c) \\
	{\rm subl.}\; 4: (a,b,c) &\mapsto (-b,-a,c). \label{eq:localrotate4}
\end{align}
\end{section}

\begin{section}{Raman scattering\label{sec:raman}}
In the search for orbitons, Raman scattering has been an important tool for
experimentalists. After the controversial first observation of orbitons in
LaMnO$_3$\cite{Saitoh2001,Grueninger2002,Saitoh2002}, the titanates now seem
to be a more promising candidate. In addition to the reasons mentioned in
previous sections, recent Raman data by Ulrich {\it et al.}\cite{Ulrich2006} 
should be noted, which shows a striking temperature
dependence: the spectral weight of the $235$ meV peak in YTiO$_3$ increases
dramatically when temperature is lowered. This can be naturally explained by
collective orbitons: as temperature drops, the orbitons gain coherence and the
spectral weight increases, analogous to two-magnon Raman scattering in the
cuprates\cite{Knoll1990}. From the local $dd$-excitation point of view,
temperature should not affect the intensity of local transitions between
crystal field levels. Also, Ulrich {\it et al.} found that the polarization
dependence of the spectra is hard to reconcile with the local excitation
picture a result that we will reproduce below. In
optical data\cite{Rueckamp2005}, a peak is seen at the same energy and was ascribed to orbital excitations.

Earlier theoretical work on Raman scattering in the
titanates\cite{Ishihara2004} built on the assumption that
JT-distortions determine the symmetry of the orbital
order. In this paper, we investigate the Raman spectrum of YTiO$_3$ in
both the lattice distortion and superexchange frameworks laid out in Sec.~\ref{sec:titanates}. We start out with the Loudon-Fleury effective Raman scattering operator\cite{Elliott1963,Fleury1968}
\begin{equation}
	\hat{R} \propto \sum_{\av{i,j}} \left( {\boldsymbol \epsilon}_i \cdot
          {\boldsymbol \delta}_{ij} \right) \left( {\boldsymbol \epsilon}_f \cdot
          {\boldsymbol \delta}_{ij} \right) \left(
          \hat{A}^{(\gamma)}_{ij} + \frac{n_{\gamma,i}+n_{\gamma,j}}{2} \right) \label{eq:Loudon-Fleury}
\end{equation}
where the usual spin exchange Hamiltonian has been replaced by the
orbital Hamiltonian of Eq.~(\ref{eq:H}). ${\boldsymbol
  \epsilon}_{i,f}$ are the electric field vectors of the in- and
out-going light, ${\boldsymbol \delta}_{ij}$ connects nearest
neighbors $i$ and $j$. The physical picture is that the light induces
an electric dipole transition to the intermediate state where a
3$d\;t_{2g}$ electron ends up on a neighboring Ti ion, after which one
of the electrons of this now doubly occupied site can hop back in
another transition. In this process, the two involved electrons can
end up in different orbitals, resulting in a two-orbiton excitation,
in full analogy with two-magnon Raman scattering in the cuprates. As the light
forces the electrons to perform a superexchange process independently
of the intrinsic coupling mechanism of the orbitals, this effective
Raman operator holds for the lattice distortion model too.

With this scattering operator, we calculate the Raman 
spectrum for the superexchange model. Similar calculations have been
done before in the context of Raman scattering on orbital excitations 
in vanadates\cite{Miyasaka2005}. Adopting the geometry used in the 
experiment of Ref.~[\onlinecite{Ulrich2006}],
we take the electric field vectors to be in the plane parallel to the $[110]$-
and $[001]$-directions: ${\boldsymbol \epsilon}_{i(f)} \propto
(\frac{1}{\sqrt{2}}\sin 
\theta_{i(f)},\frac{1}{\sqrt{2}}\sin \theta_{i(f)},\cos \theta_{i(f)})$ where
$\theta_{i(f)}$ is the angle the electric field vector makes with the
$c$-axis. Throughout this paper we use a coordinate system in which the
nearest neighbor Ti-Ti bonds are parallel to the coordinate axes. Substituting
into Eq.~(\ref{eq:Loudon-Fleury}) and using that $\sum_i n_{\gamma,i}$
is a conserved quantity in the superexchange model and that $\hat{H}_0 \ket{0} \propto
\ket{0}$, we find for inelastic Raman scattering 
\begin{equation}
	\hat{R} \propto \left( \cos \theta_i \cos \theta_f - \frac{1}{2} \sin \theta_i \sin \theta_f \right) \sum_{\av{i,j}_c} \hat{A}^{(c)}_{ij}
\end{equation}
where the sum is over bonds in the $c$-direction only. Performing the transformations mentioned in Sec.~\ref{sec:titanates}, condensing $\psi_c$ and Fourier transforming, we obtain
\begin{align}
	\sum_{\av{i,j}_c} \hat{A}^{(c)} &= \frac{2}{3} \sum_{\bf k} \left[ (a^{\dag}_{\bf k} - b^{\dag}_{\bf k})(a^{\phantom{\dag}}_{\bf k} - b^{\phantom{\dag}}_{\bf k}) + \frac{c_z}{2} (a^{\dag}_{\bf k} - b^{\dag}_{\bf k}) \times \right. \nonumber \\
	&\left. (a^{\dag}_{-{\bf k}} - b^{\dag}_{-{\bf k}}) + \frac{c_z}{2} (a^{\phantom{\dag}}_{-{\bf k}} - b^{\phantom{\dag}}_{-{\bf k}})(a^{\phantom{\dag}}_{\bf k} - b^{\phantom{\dag}}_{\bf k}) \right] \label{eq:sumAc}
\end{align}
where only quadratic terms in the operators are retained. Linear terms do not appear. Next, this result is Bogoliubov transformed according to
\begin{align}
	a_{\bf k} &= u_{\bf k} \ch \theta_{1,{\bf k}} \alpha^{\phantom{\dag}}_{1,{\bf k}} + v_{\bf k} \ch \theta_{2,{\bf k}} \alpha^{\phantom{\dag}}_{2,{\bf k}} \nonumber \\
	&- u_{\bf k} \sh \theta_{1,{\bf k}} \alpha^{\dag}_{1,-{\bf k}} - v_{\bf k} \sh \theta_{2,{\bf k}} \alpha^{\dag}_{2,-{\bf k}}, \\
	b_{\bf k} &= -v_{\bf k} \ch \theta_{1,{\bf k}} \alpha^{\phantom{\dag}}_{1,{\bf k}} + u_{\bf k} \ch \theta_{2,{\bf k}} \alpha^{\phantom{\dag}}_{2,{\bf k}} \nonumber \\
	&+v_{\bf k} \sh \theta_{1,{\bf k}} \alpha^{\dag}_{1,-{\bf k}} - u_{\bf k} \sh \theta_{2,{\bf k}} \alpha^{\dag}_{2,-{\bf k}},
\end{align}
where the indices $1,2$ refer to the orbiton branch. This transformation diagonalizes $\hat{H}_0$ up to quadratic order if
\begin{align}
	u_{\bf k} &= \sqrt{\frac{1}{2}+\frac{\gamma_{2,{\bf k}}}{2\kappa_{\bf k}}} \\
	v_{\bf k} &= {\rm sign} (\gamma_{3,{\bf k}}) \sqrt{\frac{1}{2}-\frac{\gamma_{2,{\bf k}}}{2\kappa_{\bf k}}} \\
	\tanh 2\theta_{1(2),{\bf k}} &= \gamma_{1,{\bf k}} \pm \kappa_{\bf k}.
\end{align}
The effective Raman scattering operator now either produces two orbitons or
scatters single orbitons already present in the initial state. At zero
temperature, the initial state has no orbitons (in ``linear orbital wave
theory'', i.e. if we neglect orbiton-orbiton interactions), so we keep only
the two-orbiton creation part of $\sum_{\av{i,j}_c}\hat{A}^{(c)}$ in
Eq.~(\ref{eq:sumAc}): 
\begin{align}
	\frac{1}{3} &\sum_{\bf k} \left[ \left\{ (u+v)^2 (c_z \ch 2 \theta_1 - \sh 2 \theta_1 ) \right\} \alpha^{\dag}_{1,{\bf k}} \alpha^{\dag}_{1,-{\bf k}} \right. \nonumber \\
	&+\left\{ (u-v)^2 (c_z \ch 2 \theta_2 - \sh 2 \theta_2 ) \right\} \alpha^{\dag}_{2,{\bf k}} \alpha^{\dag}_{2,-{\bf k}} \nonumber \\
	&+2\left\{ (u^2-v^2) [\sh (\theta_1 + \theta_2) \left. - c_z \ch
            (\theta_1 + \theta_2 )] \right\} \alpha^{\dag}_{1,{\bf k}}
          \alpha^{\dag}_{2,-{\bf k}} \right] 
\label{eq:matrixelem.}
\end{align}
where $c_z = \cos k_z$ and the index ${\bf k}$ is implied on every
$u,v,\theta_1$ and $\theta_2$. 

The cross section at zero temperature now is
\begin{equation}
	\frac{d^2\sigma}{d\omega d\Omega} \propto \sum_f \abs{\bra{f}\hat{R}\ket{0}}^2 \delta (\omega - \omega_f)
\end{equation}
with $f$ labelling the two-orbiton final states with energy $\omega_f$. The
corresponding matrix elements are given by Eq.~(\ref{eq:matrixelem.}).  

Because there are orbiton-orbiton interaction terms in the Hamiltonian which are neglected in ``linear orbital wave theory'', we introduce a phenomenological orbiton damping of $\gamma = 30$ meV. Also, broadening from other sources such as interaction with phonons and magnons can be mimicked this way.

The result is displayed in Fig.~\ref{fig:ramanspectrum}, compared to the data
from Ref.~[\onlinecite{Ulrich2006}]. In the superexchange model, only 
two-orbiton creation processes contribute to the Raman spectrum. The best 
fit is obtained for $J_{\rm orb} = 65$ meV, close to the value estimated in
Ref.~[\onlinecite{Khaliullin2003}] from magnon data of 
YTiO$_3$\cite{Ulrich2002}. Including orbiton-orbiton interactions will probably
reduce the peak energy (in analogy to two-magnon Raman scattering), 
increasing the fit parameter $J_{SE}$.

The local model of YTiO$_3$ also yields Raman spectra
via Eq.~(\ref{eq:Loudon-Fleury}). In this model, the orbital order
makes the $c$-direction different from the $a$ and $b$
ones. Therefore, all bond directions are considered separately. For
technical convenience, the rotations Eqs.~(\ref{eq:localrotate1})
through (\ref{eq:localrotate4}) are first performed. Bonds in the
$c$-direction connect sublattice $1$ to sublattice $3$, and $2$ to
$4$. Both these bonds give the same contribution to the Raman
operator:
\begin{align}
    \sum_{\av{i,j}_c} &\left( \hat{A}^{(c)}_{ij} + \frac{1}{2} \left(
        n_{c,i}+n_{c,j} \right) \right) = \\
\sum_{\av{i,j}_c} &\left( n_{a,i}n_{a,j}+n_{b,i}n_{b,j}+ a^{\dag}_i
    b^{\phantom{\dag}}_i b^{\dag}_j a^{\phantom{\dag}}_j + b^{\dag}_i
    a^{\phantom{\dag}}_i a^{\dag}_j b^{\phantom{\dag}}_j \right.  \nonumber \\
 &\left. + \frac{1}{2}\left( n_{a,i} + n_{b,j} \right) \right). \nonumber 
\end{align}
Note that the expression is symmetric in $i,j$. Similarly, for the
$a$- and $b$-directions, we obtain again the same contribution for both bonds with $i\in$
sublattice $1$ and $j \in$ sublattice $2$, and for bonds with $i\in 3$ and $j\in 4$:
\begin{align}
    \sum_{\av{i,j}_a} &\left( n_{b,i}n_{a,j}+n_{c,i}n_{c,j}+ b^{\dag}_i
    c^{\phantom{\dag}}_i c^{\dag}_j a^{\phantom{\dag}}_j + c^{\dag}_i
    b^{\phantom{\dag}}_i a^{\dag}_j c^{\phantom{\dag}}_j \right. \nonumber \\
    &\left. + \frac{1}{2}\left( n_{a,i} + n_{b,j} \right) \right), \\
    \sum_{\av{i,j}_b} &\left( n_{a,i}n_{b,j}+n_{c,i}n_{c,j}+ a^{\dag}_i
    c^{\phantom{\dag}}_i c^{\dag}_j b^{\phantom{\dag}}_j + c^{\dag}_i
    a^{\phantom{\dag}}_i b^{\dag}_j c^{\phantom{\dag}}_j \right. \nonumber \\
  &\left. + \frac{1}{2}\left( n_{b,i} + n_{a,j} \right) \right).
\end{align}
In general, these operators give rise to final states with one
and two $dd$-excitations. Using the local wave functions proposed in
Ref.~[\onlinecite{Mizokawa1996}], final states with one $dd$-excitation cannot be
reached in $(z,z)$ polarization configuration, in agreement with
the findings of Ref.~[\onlinecite{Ishihara2004}]. Because the wave
functions Eqs.~(\ref{eq:local1}) through (\ref{eq:local3}) of Pavarini {\it
et al.} are close to these states, there is little single $dd$-excitation 
weight (in particular in $(z,z)$ polarization), and the spectrum is dominated
by double $dd$-excitations. In the numerical calculations of the Raman
spectra, the same broadening of $\gamma = 30$ meV as above is included.

The resulting Raman spectra are shown in
Fig.~\ref{fig:ramanspectrum}, together with the experimental data. The experimental data peaks around
$230$ meV in the $(z,z)$ polarization configuration shown
here. In the experiment, other configurations give very similar line shapes, with the
maximum shifting around no more than $\sim 40$ meV. The intensity is
strongest when both in- and outgoing polarizations are directed along
one of the cubic axes\cite{Ulrich2006}, i.e., in the $zz,xx,yy$ polarization
geometries.

Even though we have included possible orbiton-orbiton 
interactions only as a phenomenological damping, the superexchange model 
gives a very good fit to the experimental line shape: it reproduces a single 
peak without internal structure at approximately the right energy. 
The cubic isotropy of the superexchange model is in agreement with 
experiment, as noted in Ref.~[\onlinecite{Ulrich2006}].

An interpretation of the Raman spectrum in terms of local crystal field
excitations is problematic. Not only is the
predicted strong polarization dependence of the intensity (a stark
contrast between the $c$-axis and the $a,b$-axes) opposite of what is seen in
experiment (which obeys cubic symmetry\cite{Ulrich2006}), 
the suppression of the single $dd$-excitations with
respect to double excitations leads to a wrong prediction of the peak
energy. We tried to include corrections to the Raman operator from nondiagonal
hoppings between $t_{2g}$ orbitals but this did not improve the fit. 
Also, to blur the multiple peaks together into one peak, a large
broadening is needed. Finally, the temperature dependence of the peak as
observed in Ref.~[\onlinecite{Ulrich2006}] is difficult to explain in the
context of local $dd$-excitations.

\begin{figure}[!htp]
	\begin{center}
	\includegraphics[width=\columnwidth]{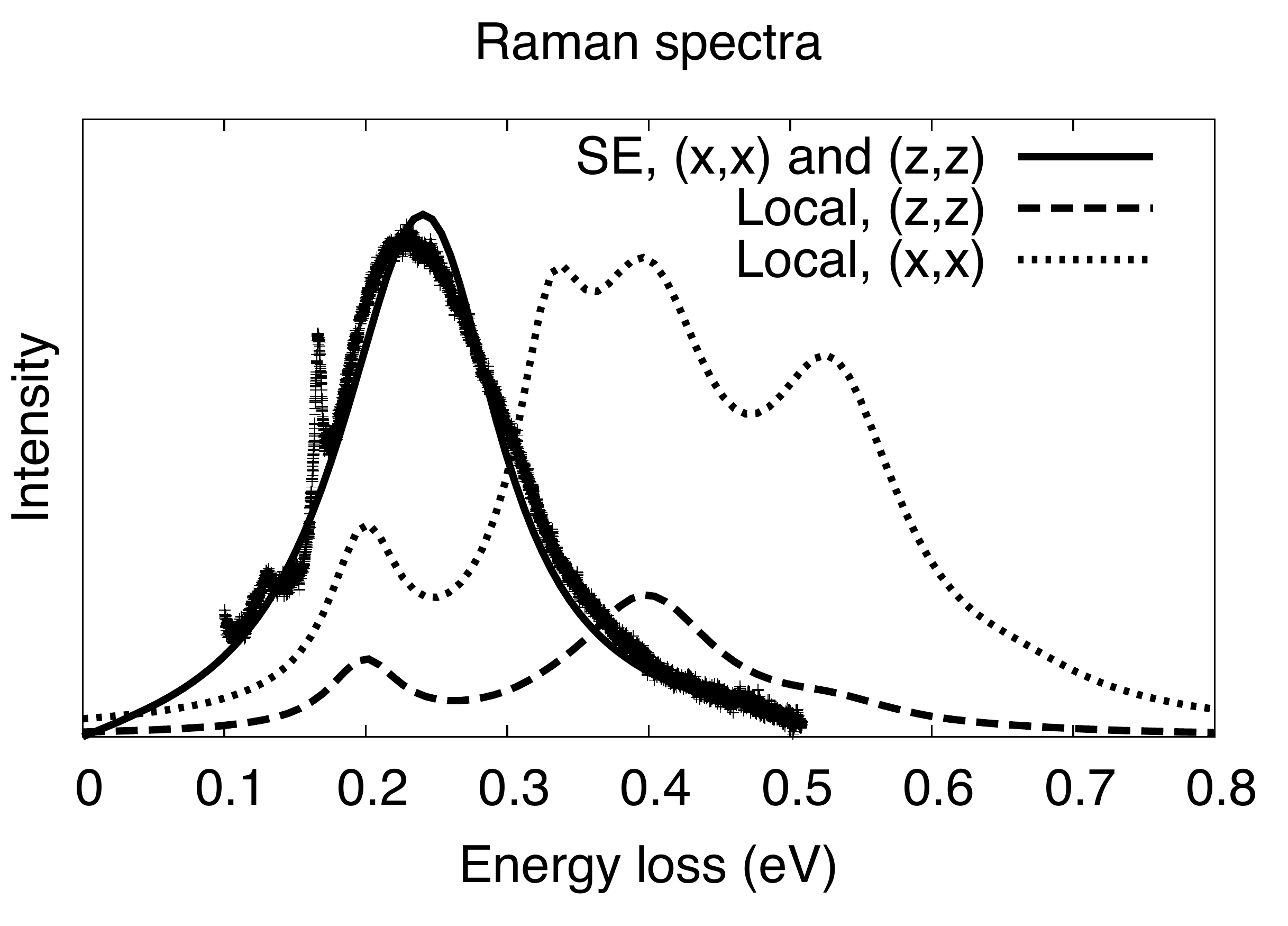}
	\caption{Raman spectrum of YTiO$_3$ at $T = 13$ K 
          in $(z,z)$ geometry, taken from Ref.~[\onlinecite{Ulrich2006}]. A
          background is subtracted from the data. The sharp peak
          around $170$ meV in the data is the two-phonon Raman signal,
          and is not considered in our theory. The
          thin-solid line is the superexchange theory curve. The
          anisotropy of the local model is reflected in its Raman
          spectra: $(z,z)$ polarization (dashed line) gives a very different
          spectrum from $(x,x)$ polarization (dotted line). In the
          superexchange model, the $xx,yy$- and $zz$-polarizations are
          equivalent. It should be noted that the experimental Raman
          spectra are
          also of cubic symmetry\cite{Ulrich2006}. \label{fig:ramanspectrum}}
	\end{center}
\end{figure}
\end{section}

\begin{section}{RIXS\label{sec:rixs}}
The rapidly developing technique of Resonant Inelastic X-ray Scattering (RIXS) is an excellent probe of collective excitations in transition metal compounds\cite{Brink2007,Hill2008,Forte2008A,Forte2008B,Braicovich2009}. The reason for this is that the X-rays carry enough momentum to map out the entire or at least a significant part of the Brillouin Zone, depending on the energy of the resonant edge used. Due to the recent advances in energy resolution, it is now possible to access energies as low as $\sim 50$ meV. This makes it in principle feasible to observe orbitons.

In RIXS, the incoming X-rays excite a core electron into or above the
valence band. Because it is a resonant technique, RIXS is
element-specific. Not only does this add more
control to the experiments, it also helps with interpreting the RIXS
spectra. In our case we consider the Ti 2$p$ to 3$d$ transition. This
excitation can in principle affect the valence electrons in two ways:
firstly through the core hole-excited electron pair's potential (from
hereon referred to simply as core hole potential) and secondly, if the
electron is excited into the valence band, by the Pauli exclusion
principle. Because of these interactions with the valence electrons,
the core hole can lose energy and momentum to the valence
electrons. The core hole can affect the valence
electrons on the core hole site itself, but it can also frustrate the
bonds of this site with its neighbors. The intermediate state is
shortlived, and when the photo-excited
electron annihilates the core hole, the energy and momentum
of the resulting X-ray photon are measured. From this measurement, it
can be deduced what the energy and momentum are of the created excitations in the
solid.

In the experiment \cite{experimentalpaper} we analyze, the $L_3$
edge is used, where the 2$p$ core electron is promoted from the
spin-orbit split $j = 3/2$ state to a 3$d$ state. The
intermediate states have a complicated multiplet structure, with large
spin-orbit coupling in the core levels, strong intra-ionic Coulomb 
interactions altered by the core potential, etc, which makes the RIXS process
hard to analyze microscopically in an exact way. 
Fortunately, it is possible to disentangle the problem of the intermediate states from the low-energy orbital 
transitions in the final states. Namely, since the intermediate states 
dynamics is much faster than that of orbital fluctuations, one can 
construct -- based on pure symmetry grounds -- a general RIXS operator 
describing orbital transitions between the initial and final states. In this
operator, the problem of the intermediate states
can be cast in the form of 
phenomenological matrix elements that depend only on the energy of the incident 
photon and its polarization factors. These martix elements can then be
calculated independently, e.g., by means of well developed quantum-chemistry 
methods on small clusters. This approach is general,
but can be simplified in the (physically relevant)
case where the energy dependence of martix elements is smooth: they
can then be regarded as effective constants at energy scales 
corresponding to the low-frequency orbital dynamics.

RIXS spectra are described  
by the Kramers-Heisenberg formula, which can be written in terms
of an effective scattering operator $\hat{O}_{\bf q}$:
\begin{equation}
	A_{fi} = \bra{f} \hat{D} \frac{1}{E_i-H-i\Gamma}\hat{D} \ket{i} = \bra{f}\hat{O}_{\bf q} \ket{i} \label{eq:KH}
\end{equation}
where $E_i$ is the incoming photon's energy, $H$ is the Hamiltonian
and $\hat{D}$ the dipole transition operator. $\Gamma$ is the lifetime
broadening of the intermediate states. The cross section is obtained via
\begin{equation}
	\frac{d^2\sigma}{d\omega d\Omega} \propto \sum_f \abs{A_{fi}}^2 \delta (\omega - \omega_{fi}).
\end{equation}
Here $\omega_{fi}$ is the energy difference between the final and
initial state of the solid. The cross section can also be written in terms of the Green's function for the effective scattering operator:
\begin{equation}
	\frac{d^2\sigma}{d\omega d\Omega} \propto \sum_f \abs{\bra{f} \hat{O}_{\bf q} \ket{i}}^2 \delta (\omega-\omega_{fi}) = -\frac{1}{\pi} \mathfrak{Im} \left\{ G(\omega) \right\} \label{eq:Greensfunction1}
\end{equation}
with
\begin{equation}
	G(\omega) = -i \int_{0}^{\infty} dt e^{i \omega t} \bra{i} \hat{O}^{\dag}_{\bf q}(t) \hat{O}^{\phantom{\dag}}_{\bf q}(0) \ket{i}. \label{eq:Greensfunction2}
\end{equation}

The effective scattering operator can in general be 
expanded in the number of sites involved in the scattering process:
\begin{equation}
  \hat{O}_{\bf q} = \sum_i e^{i{\bf q}\cdot {\bf R}_i} \left(
    \hat{O}_i + \hat{O}_{ij} + \dots \right) \label{eq:Oexpansion}
\end{equation}
where ${\bf q}$ is the transfered momentum. The phase factor comes from the dipole operators.
We neglect RIXS processes that create excitations on more than two
sites in the final state, and further
assume that the two-site processes are dominated by processes on
nearest neighbors.

One may distinguish two regimes for RIXS processes: in the first regime
$\Gamma$ is much larger than the relevant energy scales of the
intermediate states, and these processes can be easily analyzed with the
Ultrashort Core hole Lifetime expansion\cite{Brink2005,Ament2007}. In
the other regime $\Gamma$ is small and its inverse is irrelevant as a cut-off
time of the intermediate state dynamics. The lifetime broadening at
the transition metal $L$ edges is relatively small, and the
effects of the core hole on the valence electrons is averaged over
many precessions of the core hole due to the large spin-orbit coupling 
in the core levels of transition metal ions. Therefore the $A_{1g}$ component 
of the Coulomb potential of the core hole dominates the scattering processes. 
In the following, we assume that the titanates belong to the regime of 
small $\Gamma$, and that the internal dynamics of intermediate states 
is the fastest process in the problem.

Returning to the scattering operator, Eq.~(\ref{eq:Oexpansion}),
we are left with two interesting cases. The single-site operator is
dominated by the $A_{1g}$ component, but this only gives
contributions to the Bragg peaks. The subleading order therefore
consists of single site processes $\hat{O}_i$ of other 
than $A_{1g}$ symmetries, and of
two-site processes $\hat{O}_{ij}$ of $A_{1g}$ symmetry.

The single site coupling of RIXS to the orbitals can be dubbed a ``shakeup'' process. If we
allow the core hole potential to have a symmetry other than $A_{1g}$, it can
locally induce an orbital flip. If the orbital ground-state is dominated by
superexchange many-body interactions, a local flipped orbital will 
strongly interact with the neighboring sites and thus becomes a superposition 
of extended (multi-)orbitons. In the limit of strong crystal field splittings, 
however, this excitation remains a localized, on-site transition between 
$t_{2g}$ levels.

Two-site processes $\hat{O}_{ij}$ may involve
modulation of the superexchange bonds, analogous to two-magnon
RIXS, where the superexchange constant $J$ is
effectively modified at the core hole
site\cite{Brink2007,Forte2008A,Forte2008B,Braicovich2009}. 
The core hole potential locally changes the Hubbard $U$, which
in effect changes $J_{SE} = 4 t^2/U$ on the Ti-Ti bonds coupled to the core hole
site. Alternatively, the two-site processes can describe the
lattice-mediated interaction that is altered by the presence of a core
hole. The equilibrium positions and vibration frequencies of the
oxygens surrounding the core hole site may change, affecting the intersite
interactions. As said above, the $A_{1g}$ component of the
core hole potential is most relevant in the two-site coupling
channel $\hat{O}_{ij}$.

This section is divided into three subsections. Subsection \ref{subsec:shakeup}
deals with the single site shakeup mechanism and contains the
evaluation in the superexchange model. The next subsection,
\ref{subsec:local}, is devoted to the calculation of the same
processes in the local model of the orbital
excitations in YTiO$_3$. The final subsection
\ref{subsec:SE} covers two-site processes, evaluated within
the superexchange model. A detailed comparison is made of the RIXS spectra arising from the different
models.

\begin{subsection}{Single site processes -- Superexchange model\label{subsec:shakeup}}
We start out with an analysis of the single site
processes. RIXS processes that involve orbital excitations on a
single site are
dominated by direct transitions between the $t_{2g}$ orbitals when the
core hole potential is not of $A_{1g}$ symmetry. In a superexchange
dominated system, a local flipped orbital 
strongly interacts with the neighboring sites
and becomes a superposition of extended orbitons.

We start from the Kramers-Heisenberg equation
\begin{equation}
	\hat{O}_{\bf q} = \hat{D} \frac{1}{E_i-\hat{H}-i\Gamma} \hat{D}.
\end{equation}
We insert the polarization-dependent dipole operator $\hat{D}$ which we take to be local: $\hat{D} = \sum_i \hat{D}_i$ with
\begin{align}
	\hat{D}_i = &\sum_{d,m} \left( e^{-i{\bf q}_{\rm in}\cdot {\bf R}_i} \ket{m}\bra{m}\hat{{\bf r}}\cdot {\boldsymbol \epsilon} \ket{d}\bra{d} \right. \nonumber \\
	&\left. + e^{i{\bf q}_{\rm out}\cdot {\bf R}_i} \ket{d}\bra{d} \hat{{\bf r}}\cdot {\boldsymbol \epsilon}' \ket{m}\bra{m} \right) + {\rm h.c.},
\end{align}
where ${\boldsymbol \epsilon}$ and ${\boldsymbol \epsilon}'$ are the in- and outgoing polarization vectors respectively, $\ket{d}$ denotes the state of atom $i$ when it is not photo-excited and $\ket{m}$ denotes the system's intermediate eigenstates:
\begin{equation}
	\hat{H} = \sum_m E_m \ket{m}\bra{m}.
\end{equation}
Now we consider only the single site part of the effective
scattering operator in Eq.~(\ref{eq:Oexpansion}):
\begin{equation}
	\hat{O}_i = \sum_{d,d',m} \ket{d'}\bra{d'}\hat{{\bf r}}' \cdot
        {\boldsymbol \epsilon}' \ket{m}
        \frac{1}{E_i-E_m-i\Gamma} \bra{m} {\boldsymbol \epsilon} \cdot \hat{{\bf r}} \ket{d} \bra{d} \label{eq:shakeupOq}
\end{equation}
Next we decompose the operator part into terms transforming according to the rows of the irreducible representations of the octahedral group (labeled by $\Gamma$, not to be confused with the core hole lifetime broadening):
\begin{equation}
	\ket{d'}\bra{d} = \sum_{\Gamma} \Gamma_{d'd} \hat{\Gamma}. \label{eq:decomposeoperator}
\end{equation}
In the second quantized picture, we need only terms that are quadratic in the creation and annihilation operators. With the irreducible representations $A_{1g}, T_{1u}, E_g$ and $T_{2g}$ all possible $\ket{d'}\bra{d}$ can be constructed. Therefore $\hat{\Gamma}$ assumes only the following forms:
\begin{align}
	A_{1g}&: \hat{\Gamma} = \openone \\
	T_{1u}&: \hat{\Gamma} \in \{ \hat{l}_x,\hat{l}_y,\hat{l}_z,\} \label{eq:T1u} \\
	E_g&: \hat{\Gamma} \in \{ \hat{Q}_x, \hat{Q}_z \} \\
	T_{2g}&: \hat{\Gamma} \in \{ \hat{T}_x, \hat{T}_y, \hat{T}_z \}. \label{eq:T2g}
\end{align}
The operators $\hat{\Gamma}$ and the corresponding $3\times 3$ matrices
$\Gamma_{d'd}$ are defined in Appendix~\ref{app:shakeup}. Because $A_{1g}$
only contributes to elastic scattering, we drop it from hereon. 

Further, we also decompose the dipole matrix elements into
\begin{equation}
	\bra{d'} \hat{\beta} \ket{m}\bra{m}\hat{\alpha} \ket{d} = \sum_{\Gamma} \Gamma_{\beta\alpha} M^{\Gamma}_{d'd} \label{eq:decomposepolarization}
\end{equation}
with $\hat{\alpha},\hat{\beta} \in \{\hat{x},\hat{y},\hat{z}\}$ and the $M^{\Gamma}_{d'd}$ listed in Appendix~\ref{app:multipletfactors}: Eqs.~(\ref{eq:MA1g}) through (\ref{eq:Mlz}). Plugging Eqs.~(\ref{eq:decomposeoperator}) and (\ref{eq:decomposepolarization}) into Eq.~(\ref{eq:shakeupOq}), we obtain
\begin{equation}
  \hat{O}_i = \sum_{d,d',m} \sum_{\Gamma'} \sum_{\alpha,\beta}
  \frac{\epsilon'_{\beta} \epsilon_{\alpha} \Gamma'_{\beta\alpha} M^{\Gamma'}_{d'd}}{E_i-E_m-i\Gamma} \sum_{\Gamma} \Gamma_{d'd} \hat{\Gamma}_i
\end{equation}
which can be simplified using
\begin{equation}
	\sum_{d,d'} M^{\Gamma'}_{d'd} \Gamma_{d'd} = \delta_{\Gamma,\Gamma'} \sum_{d,d'} M^{\Gamma}_{d'd} \Gamma_{d'd}.
\end{equation}
This identity can be proven by interpreting $M^{\Gamma}$ and $\Gamma$ as matrices indexed by $d$ and $d'$. Then it can be seen that $M^{\Gamma} \propto \Gamma$. We thus obtain
\begin{equation}
	\sum_{d,d'} M^{\Gamma'}_{d'd} \Gamma_{d'd} = \Tr{M^{\Gamma'} \Gamma^T} \propto \Tr{\Gamma' \Gamma^T}
\end{equation}
which is zero for $\Gamma \neq \Gamma'$, proving the above identity. We find then
\begin{equation}
	\hat{O}_i = \sum_{\Gamma} P_{\Gamma} \mathcal{M}_{\Gamma}
        \hat{\Gamma}_i 
\label{eq:Oi}
\end{equation}
with a polarization factor
\begin{equation}
	P_{\Gamma} = \sum_{\alpha,\beta} \epsilon'_{\beta} \Gamma_{\beta\alpha} \epsilon_{\alpha}
\end{equation}
and the matrix elements $\mathcal{M}_{\Gamma}$ depending on the multiplet
effects in the intermediate state 
\begin{equation}
  \mathcal{M}_{\Gamma} = \sum_{d,d',m}
  \frac{M^{\Gamma}_{d'd} \Gamma_{d'd}}{E_i-E_m-i\Gamma}\ . \label{eq:Mfirstorder}
\end{equation}

One can perform the sum over $m$, which yields
\begin{align}
  \sum_m &\frac{M^{Q_x}_{d'd}}{E_i-E_m-i\Gamma} = \nonumber \\
  &\bra{d'} \left( \hat{y} \frac{1}{E_i-H-i\Gamma} \hat{y} - \hat{x} \frac{1}{E_i-H-i\Gamma} \hat{x} \right) \ket{d}
\end{align}
and similar expressions for the other representations. As discussed above, we
will assume that the intermediate state dynamics is much faster than that of 
$t_{2g}$ orbitals we are interested in, and thus regard the matrix elements as
phenomenological constants. Further, using that $\mathcal{M}_{\Gamma}$ does
not depend on any coordinate and therefore must be invariant under the
octahedral group, we obtain 
\begin{align}
	\mathcal{M}_Q &\equiv \mathcal{M}_{Q_x} = \mathcal{M}_{Q_z} \\
	\mathcal{M}_T &\equiv \mathcal{M}_{T_x} = \mathcal{M}_{T_y} = \mathcal{M}_{T_z} \\
	\mathcal{M}_l &\equiv \mathcal{M}_{l_x} = \mathcal{M}_{l_y} = \mathcal{M}_{l_z}.
\end{align}
The $\mathcal{M}_{\Gamma}$ are hard to calculate explicitly since they
involve inverting $H$, which contains the multiplet
structure. In the following, we assume $\mathcal{M}_{\Gamma} =
\mathcal{M}$ for all $\Gamma$.
This is a reasonable assumption: the core hole generates a multitude
of many-body states that 
evolves very rapidly due to the large spin-orbit coupling and intra-ionic 
Coulomb interactions, and therefore
its potential is averaged. Any particular symmetry is washed
away; all become equal, except for the $A_{1g}$ component, which is enhanced at the cost of the
others. This is also the reason why the experiments at the $t_{2g}$
and $e_g$ edges are similar\cite{experimentalpaper}: the different 
edges create different multiplet structures initially, 
but these differences are averaged out by the
intermediate state dynamics, as far as we are concerned with $t_{2g}$ orbital transitions 
at relatively low energies 0.2-0.3 eV.

Note that $P_{\Gamma}$ and $\mathcal{M}_{\Gamma}$ are independent of the site $i$. Only $\hat{\Gamma}$ depends on $i$, giving
\begin{equation}
	\hat{O}_{\bf q} = \sum_{\Gamma} P_{\Gamma}
        \mathcal{M}_{\Gamma} \sum_i e^{i {\bf q}\cdot {\bf R}_i}
        \hat{\Gamma}_i. \label{eq:singlesiteO}
\end{equation}

Most interference terms between different $\Gamma$'s are zero. This comes
about because of the specific ground state ordering. Transforming to the local
axes (Eq.~(15) in Ref.~\onlinecite{Khaliullin2003}), the ground state and
$\hat{H}_0$ are invariant under translations, while the operators
$\hat{T}_{\alpha,i}$ and $\hat{l}_{\alpha,i}$ (with $\alpha \in \{x,y,z\}$)
acquire a phase upon translation to a different sublattice, which is
equivalent to a momentum shift (by orbital ordering vectors)
for the corresponding $\hat{\Gamma}_{\bf q}$. Therefore, many interference terms are zero, which can be seen from
Eqs.~(\ref{eq:Greensfunction1}) and (\ref{eq:Greensfunction2}): two operators
with different momenta cannot bring the ground state (zero momentum) back to
itself. The only non-vanishing interference terms are
$\bra{0}\hat{Q}^{\dag}_{x,{\bf q}}(t) \hat{Q}^{\phantom{\dag}}_{z,{\bf q}}(0)
\ket{0}$ which do not acquire momentum shifts and
$\bra{0}\hat{T}^{\dag}_{\alpha,{\bf q}} (t)
\hat{l}^{\phantom{\dag}}_{\alpha,{\bf q}}(0) \ket{0}$ where the momentum
shifts cancel. 

To compare with experiment, we calculate the polarization factors $P_{\Gamma}$ for the experimental setup of Ref. \onlinecite{experimentalpaper}, where ${\bf q}$ is along the $[001]$-direction. Only the incoming polarization is fixed, the outgoing polarization is not detected and should be averaged over. We have
\begin{align}
	{\boldsymbol \epsilon} &= (\frac{1}{\sqrt{2}} \sin \theta , \frac{1}{\sqrt{2}} \sin \theta , \cos \theta ) \\
	{\boldsymbol \epsilon}'_H &= (-\frac{1}{\sqrt{2}} \sin \theta , -\frac{1}{\sqrt{2}} \sin \theta , \cos \theta ) \\
	{\boldsymbol \epsilon}'_V &= (\frac{1}{\sqrt{2}}, \frac{1}{\sqrt{2}}, 0)
\end{align}
with $2\theta$ the scattering angle. Then, we find for the horizontal outgoing polarization ${\boldsymbol \epsilon}'_H$ (i.e. the electric field vector is in the scattering plane):
\begin{align}
	P_{Q_x,H} &= P_{T_x,H} = P_{T_y,H} = P_{l_z,H} = 0 \label{eq:pol_start} \\
	P_{A_{1g},H} &= \frac{1}{3}\cos 2 \theta \\
	P_{Q_z,H} &= \frac{1}{2\sqrt{3}}(1+\cos^2 \theta) \\
	P_{T_z,H} &= \frac{1}{2} \sin^2 \theta \\
	P_{l_x,H} &= - P_{l_y,H} = -\frac{i}{\sqrt{2}} \sin \theta \cos \theta
\end{align}
and for vertical outgoing polarization ${\boldsymbol \epsilon}'_V$ (electric field vector perpendicular to the scattering plane):
\begin{align}
	P_{A_{1g},V} &= P_{Q_z,V} = P_{T_z,V} = 0 \\
	P_{Q_x,V} &= -\frac{1}{2} \sin \theta \\
	P_{T_x,V} &= -P_{T_y,V} = \frac{1}{2\sqrt{2}} \cos \theta \\
	P_{l_x,V} &= P_{l_y,V} = -\frac{i}{2\sqrt{2}} \cos \theta \\
	P_{l_z,V} &= \frac{i}{2} \sin \theta. \label{eq:pol_end}
\end{align}
For horizontal polarization, the polarization factors make all remaining interference terms zero.

In Appendix~\ref{app:Oqorbitons}, the one- and two-orbiton parts of
the $\hat{\Gamma}_{\bf q} = \sum_i e^{i {\bf q}\cdot {\bf R}_i}
\hat{\Gamma}_i$ are listed. They are obtained by performing the
transformations on the orbital operators mentioned in
Ref.~[\onlinecite{Khaliullin2003}]. Then, the $\psi_c$ orbital (with corresponding annihilation operator $\tilde{c}$) is condensed:
\begin{equation}
	n_{\tilde{c}} = \abs{c_0}^2 + \delta n_{\tilde{c}}\label{eq:condense}
\end{equation}
where $\delta n_{\tilde{c}}$ is the fluctuating part. In the completely ordered state, $\av{\delta n_{\tilde{c}}} = \av{n_{\tilde{a}}} = \av{n_{\tilde{b}}} = 0$ and $\abs{c_0}^2 = 1$, while in the completely disordered state $\av{\delta n_{\tilde{c}}} = \av{n_{\tilde{a}}} = \av{n_{\tilde{b}}} = 1/3$ and $\abs{c_0}^2 = 0$. Ref.~[\onlinecite{Khaliullin2003}] obtains a finite value for the quadrupole orbital order parameter:
\begin{equation}
	\hat{Q} = n_{\tilde{c}} - (n_{\tilde{a}}+n_{\tilde{b}})/2 \equiv
        \av{\hat{Q}} + \delta \hat{Q} \simeq 0.19 + \delta \hat{Q} 
\end{equation}
with the fluctuating part averaging to zero. This fixes $\abs{c_0}^2 \simeq  
0.19$. Taking the square root of Eq.~(\ref{eq:condense}), one arrives at 
\begin{equation}
	\tilde{c} = \tilde{c}^{\dag} = \sqrt{ \abs{c_0}^2 + \delta n_{\tilde{c}}} \approx \abs{c_0} + \frac{1}{2\abs{c_0}} \delta n_{\tilde{c}}
\end{equation}
to first order in the fluctuations $\delta n_{\tilde{c}}$.

In the process of writing the $\hat{\Gamma}_{\bf q}$ in terms of orbiton
operators, unphysical contributions to the intensity may appear as a result of
neglecting cubic and higher order terms in the orbiton operators. When
restoring all terms, these unphysical contributions should cancel by
symmetry. For ${\bf q}$ along the $[001]$-direction for instance,
$[\hat{H},\hat{Q}_{z,{\bf q}}] = 0$ if we use the untransformed forms
Eqs.~(\ref{eq:H}) and (\ref{eq:Qzbare}), and it is clear that there should
only be an elastic contribution to the intensity. However, in terms of
orbitons, this selection rule is violated if we go only up to quadratic
orbiton terms. To make sure these unphysical contributions are dropped, we
first calculate the commutator in the untransformed picture. If this yields
zero, the commuting part of the scattering operator is dropped. Applying this
procedure to the case where ${\bf q}$ is along the $[001]$-direction, we find
that only $\hat{Q}_{z,{\bf q}}$ among the operators (\ref{eq:T1u}) to
(\ref{eq:T2g}) is zero while all the other channels give finite
contributions. 

Since we did not include explicitly the orbiton-orbiton interactions,
damping of the orbitons should still be taken care of, at least on a
phenomenological level. As in the case of Raman scattering calculations, 
we introduce by hand an energy broadening $\gamma$ of the orbiton states 
(half-width at half maximum, HWHM) of $\gamma = 0.4\;J_{\rm orb}$. This 
broadening can also be used to take orbiton damping by phonons, magnons etc. 
into account. In addition to this, there is an experimental broadening added of
$27.5$ meV (HWHM)\cite{experimentalpaper}. 

The resulting spectra are shown in Fig.~\ref{fig:shakeup}. The intensity is 
strongly momentum-dependent (especially for ${\bf q}$ along the 
$z$-direction), which is also seen in the experiments\cite{experimentalpaper}. 
This dependence is mainly due to the coherent response of the 
exchange-coupled orbitals which enhances at large momenta, reflecting
staggered orbital order in the ground state -- Eq.~(\ref{eq:psiSE}).    
In Fig.~\ref{fig:shakeupcompare}, the
theoretical cross section (with ${\bf q}$ along the $z$-direction and
horizontal incoming polarization, i.e. the electric field is in the scattering
plane) is compared to the experimental data. The main features of the
data\cite{experimentalpaper} are reproduced: the spectral weight increases
with increasing $q_z$ and there is virtually no dispersion of the maximum of
the theoretical curve (because it is determined by the two-orbiton
continuum, containing an integration over the Brillouin zone).

Especially in the second and third plots, the
one-orbiton shoulder seems a bit too large. However, we note that there are
several factors that can alter the line shape. First we note again that the weight of this shoulder is controlled by the
orbital order parameter: if the orbital order melts, $\abs{c_0}^2$ decreases
and the one-orbiton peak becomes less intense. The value we used ($\abs{c_0}^2
= 0.19$) is obtained at zero temperature, assuming that YTiO$_3$ is a fully
saturated ferromagnet\cite{Khaliullin2002,Khaliullin2003}. Under realistic
conditions, $\abs{c_0}^2$ is expected to be smaller than $0.19$. Indeed, the
saturated ordered moment in YTiO$_3$ is actually $0.84 \mu_B$, which is
reduced further to approximately $0.80 \mu_B$ at $T = 15$
K\cite{Garrett1981,Goral1982}. Correspondingly, the orbital order is decreased
by joint spin-orbital quantum fluctuations, suppressing the one-orbiton peak. 

Secondly, we assumed all $\mathcal{M}_{\Gamma}$ are equal. Different values
would correspond to different line shapes. We note that the $T_{2g}$
representation has a much reduced one-orbiton contribution compared to the
other channels. 

Thirdly, we introduced the finite orbiton lifetime broadening as a
phenomenological damping only. All vertex corrections
to the two-orbiton diagram are neglected. In analogy to two-magnons, these
terms can give corrections to the spectrum.

The best chance to see a one-orbiton contribution to the spectrum 
is with momentum transfer directed maximally in the
$[110]$-direction. Fig.~\ref{fig:shakeup110} shows the prediction for the
shakeup mechanism with $\abs{c_0}^2 = 0.19$: the one-orbiton peak is about as
strong as the two-orbiton peak. 

\begin{figure}[!htp]
	\begin{center}
	\includegraphics[width=.75\columnwidth]{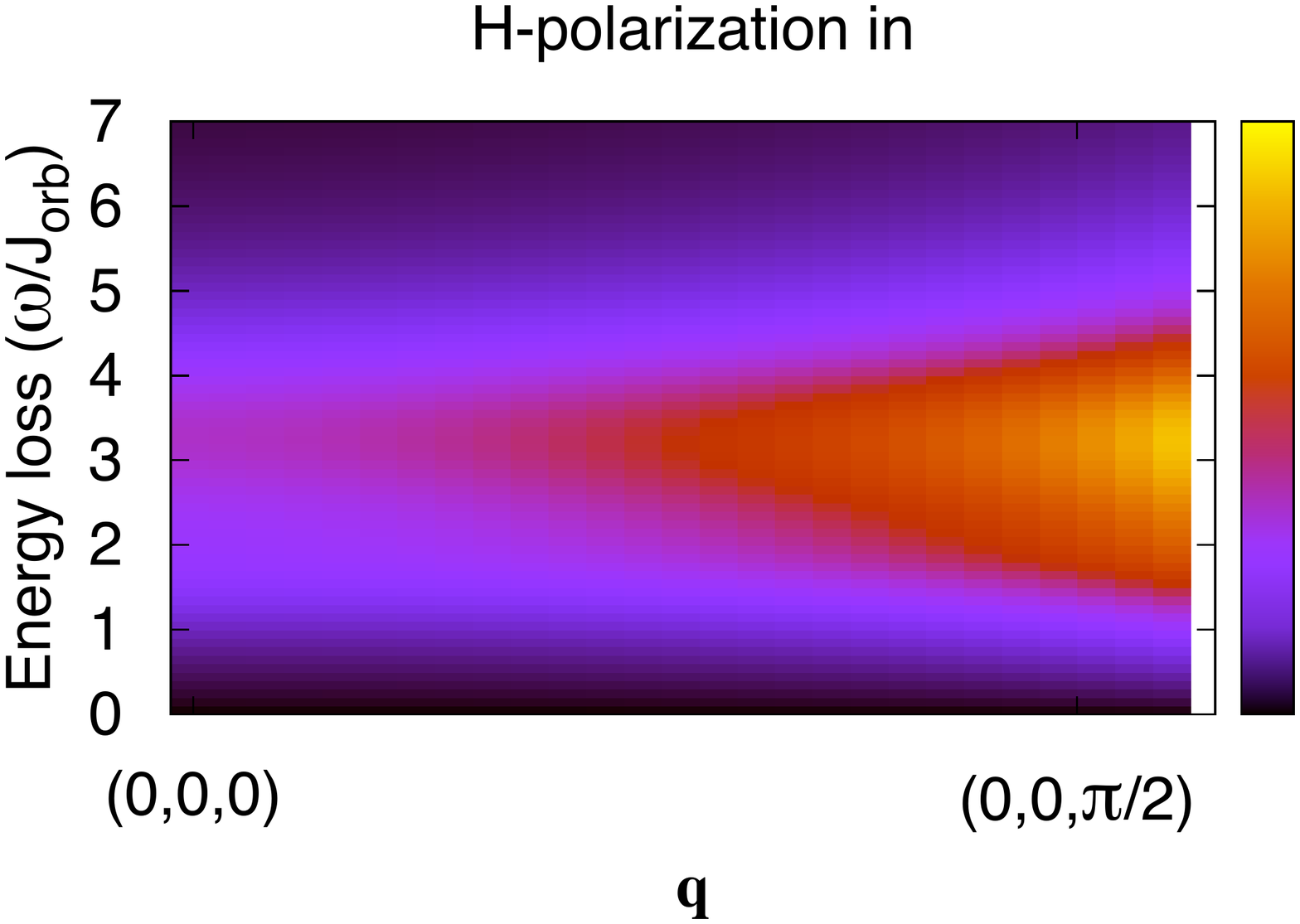}
	\includegraphics[width=.75\columnwidth]{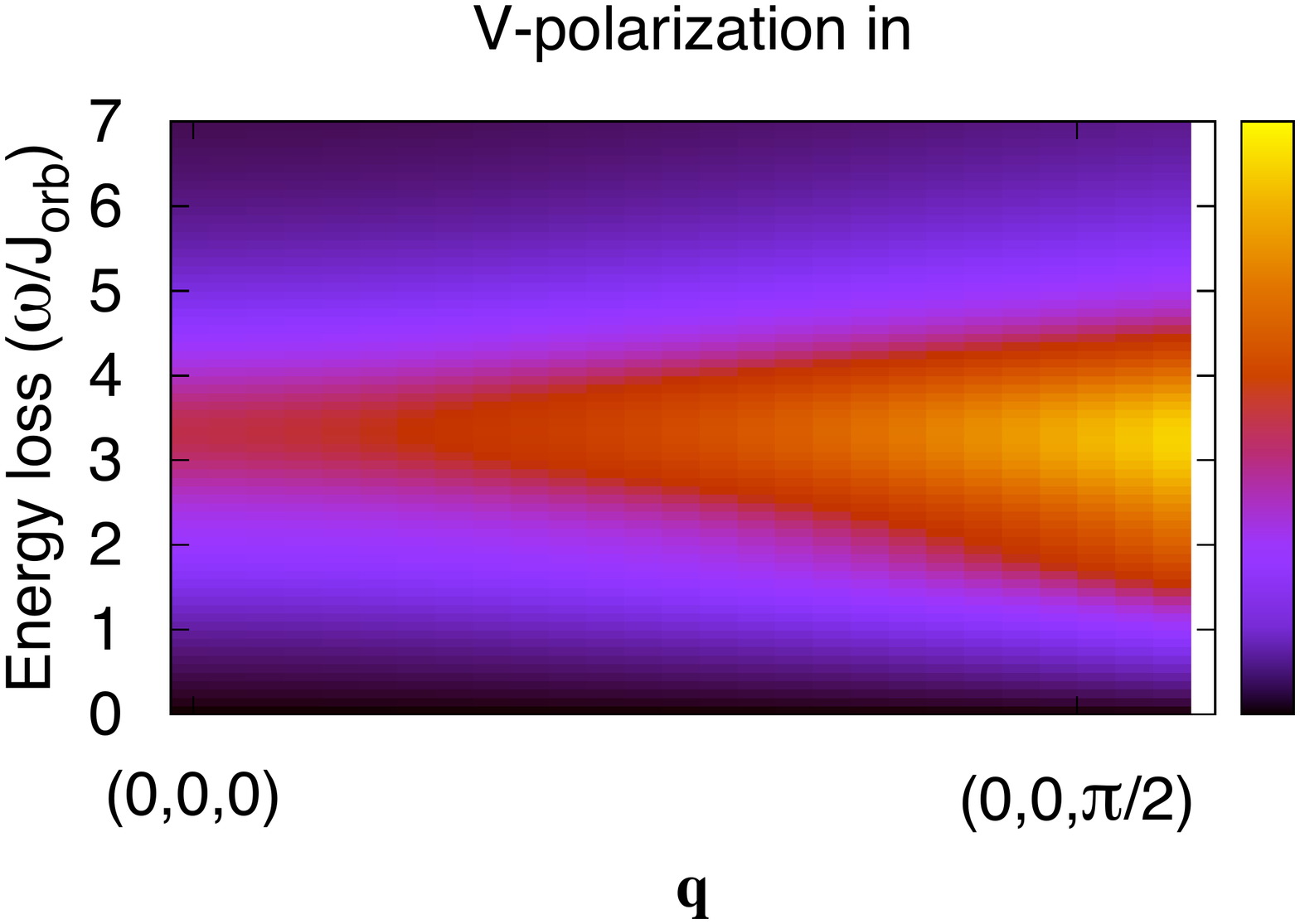}
	\includegraphics[width=.75\columnwidth]{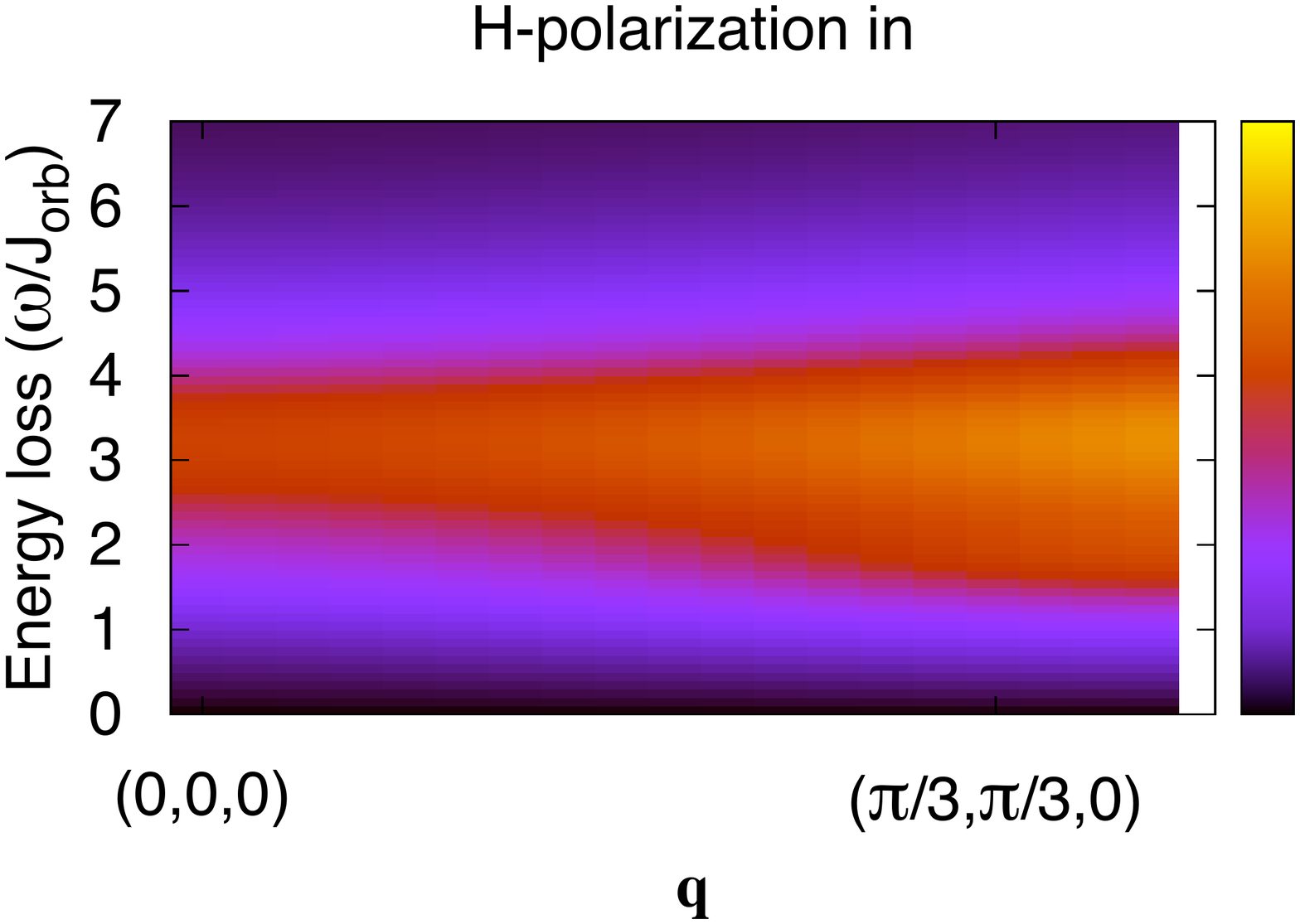}
	\includegraphics[width=.75\columnwidth]{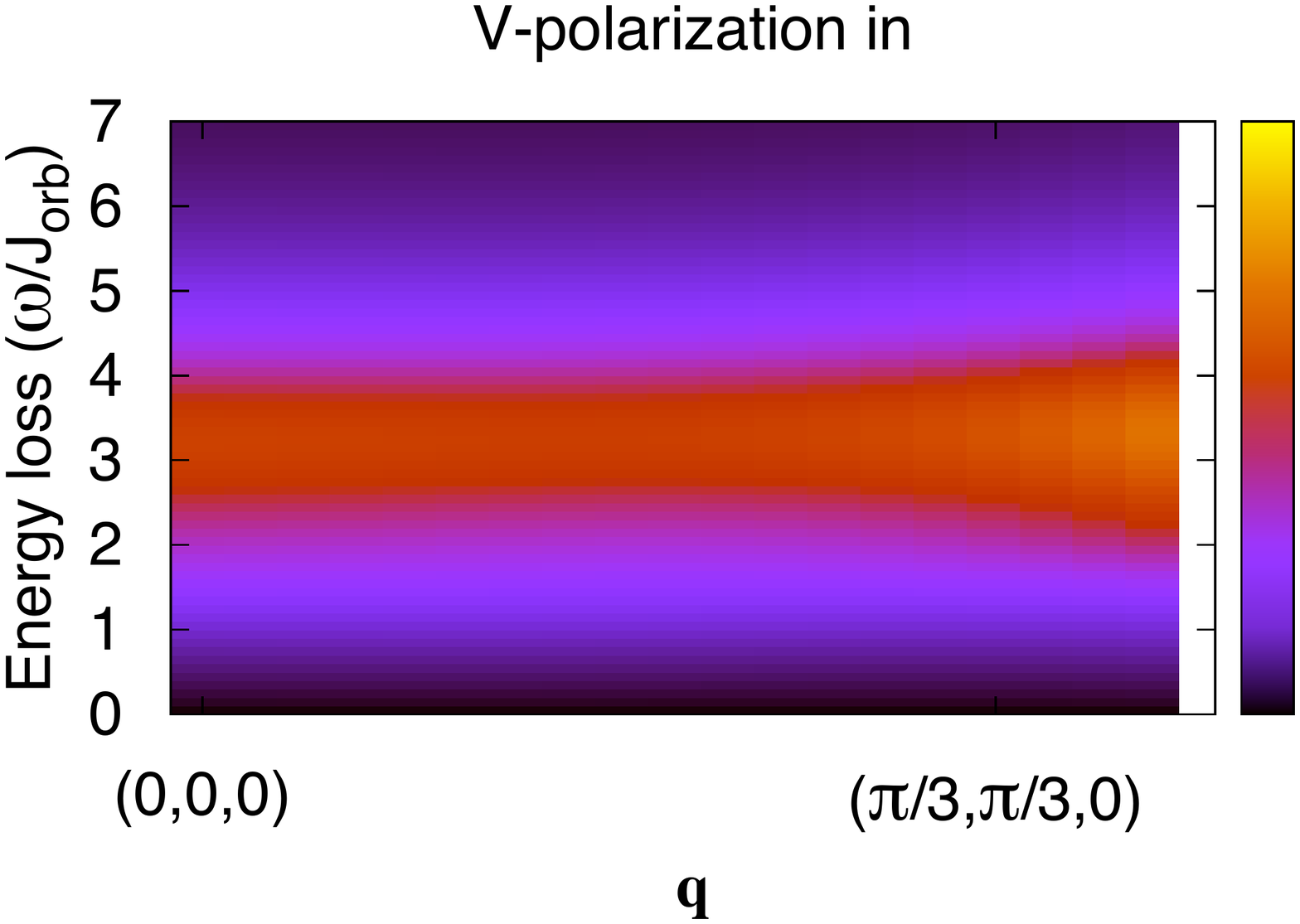}
	\caption{(Color online) RIXS spectra for a model of superexchange-driven
          orbital order with RIXS coupling to orbitons only via the
          single site mechanism. The first two spectra are for ${\bf q}$ directed along
          the $[001]$-direction, the last two for ${\bf q}$ along the
          $[110]$-direction. The first and third spectrum are for horizontal
          incoming polarization (electric field in the scattering plane), the
          second and fourth are for vertical incoming polarization (electric
          field perpendicular to the scattering plane). Note that the ${\bf q}
          = {\bf 0}$ points are different in each spectrum because of the
          different experimental geometries, leading to different
          $P_{\Gamma}$. We only plotted the experimentally accessible part of
          the Brillouin Zone.\label{fig:shakeup}} 
	\end{center}
\end{figure}

\begin{figure}[!htp]
	\begin{center}
	\includegraphics[width=\columnwidth]{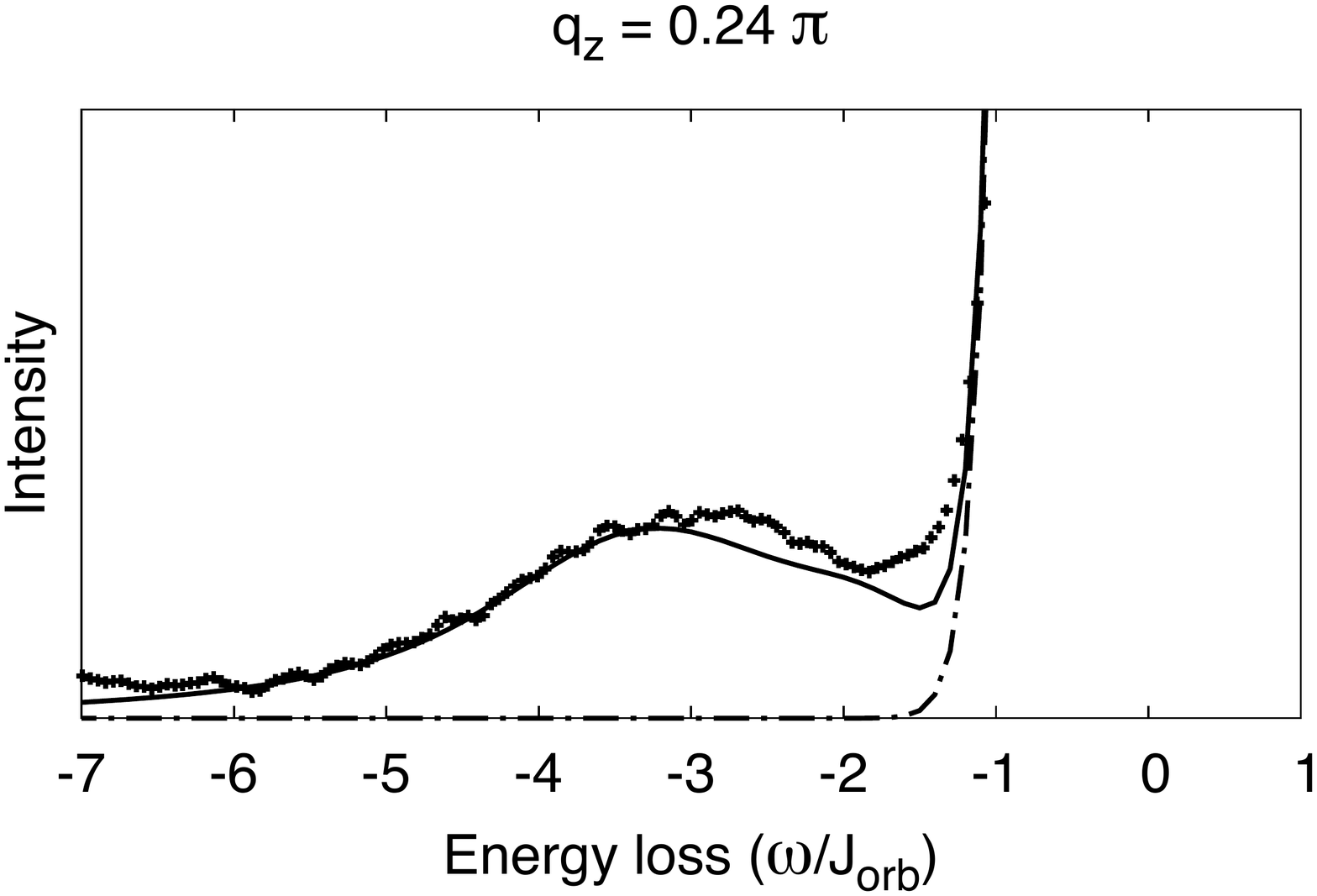}
	\includegraphics[width=\columnwidth]{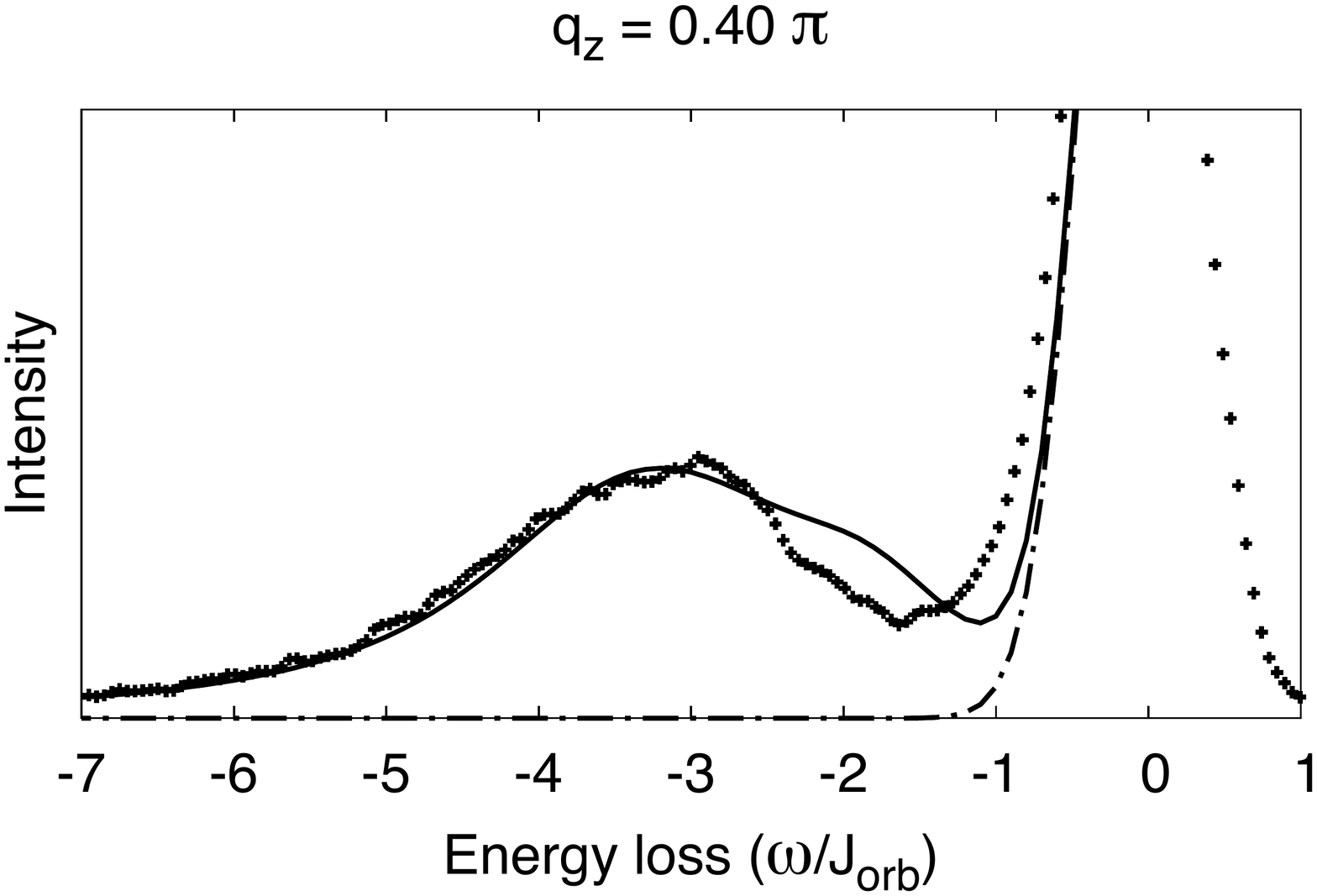}
	\includegraphics[width=\columnwidth]{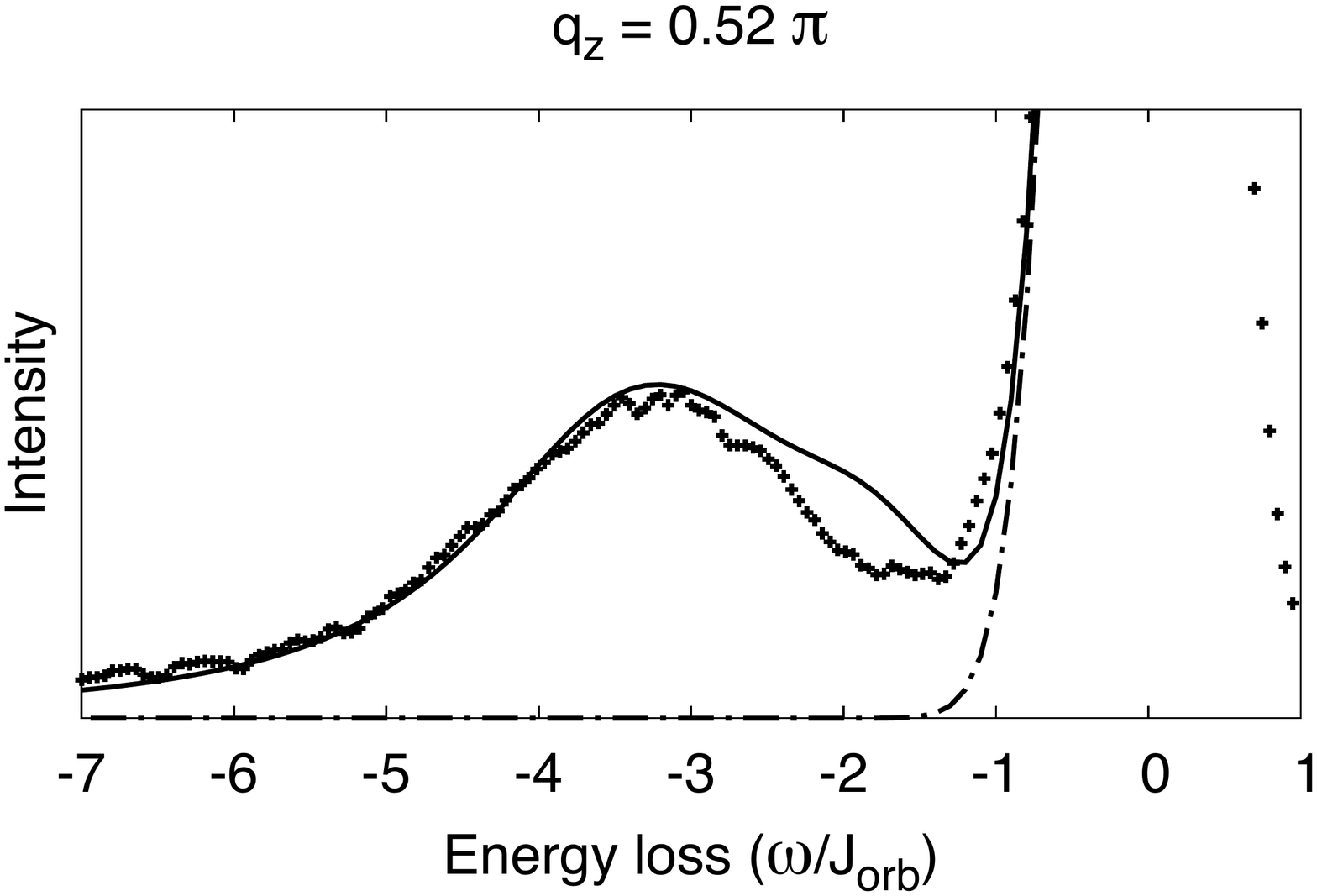}
	\caption{RIXS spectra for a superexchange-driven
          orbital order with RIXS coupling to orbitons only via the
          single site mechanism (solid line), compared to the experimental data\cite{experimentalpaper}. The vector ${\bf q}$ is directed along the $[001]$-direction, with $q_z$ as indicated in the figures. We took $J_{\rm orb} = 80$ meV, and introduced a phenomenological HWHM broadening of $\gamma = 0.4\; J_{\rm orb} \approx 30$ meV for the orbitons, as well as the HWHM experimental resolution of $27.5$ meV. The elastic peak is fitted with a Gaussian (dash-dotted line).\label{fig:shakeupcompare}}
	\end{center}
\end{figure}

\begin{figure}[!htp]
	\begin{center}
	\includegraphics[width=\columnwidth]{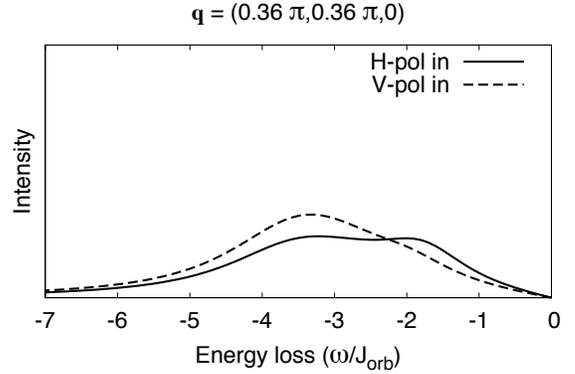}
	\caption{Spectra, obtained with the single site mechanism, for the largest experimentally accessible momentum transfer directed along the $[110]$-direction. The solid line indicates the case where the incoming polarization is horizontal, the dashed line is for vertical incoming polarization. The elastic peak has been removed. In the horizontally polarized case, the single orbiton peak is quite strong and should be visible in experiments if the system is superexchange-driven and the RIXS signal is dominated by the single site mechanism.\label{fig:shakeup110}}
	\end{center}
\end{figure}
\end{subsection}

\begin{subsection}{Single site processes -- Local model\label{subsec:local}}
Although the response functions of the local model of
  YTiO$_3$ are entirely different from the superexchange model, the
  phenomenological scattering operator Eq.~(\ref{eq:Oexpansion}) is
  still valid. Focusing on single site processes, Eq.~(\ref{eq:singlesiteO})
  can be evaluated using the wave functions found by Pavarini {\it et al.}:
  Eqs.~(\ref{eq:local1}) through (\ref{eq:local3}). Since the
eigenstates of the local model have a very simple form, we can
straightforwardly use Eqs.~(\ref{eq:lxbare}) through (\ref{eq:Tzbare})
to evaluate the RIXS spectrum. The $P_{\Gamma}$ and
$\mathcal{M}_{\Gamma}$ remain the same as in the collective orbiton
case. The spectrum now consists of two sharp peaks at $\omega_1$ and
$\omega_2$. These peaks can be broadened by coupling to the lattice as
well as due to the superexchange coupling.

Because there are four sublattices which all support their own, local
eigenstates, the RIXS intensity can be decomposed into four
signals. From the expressions Eqs.~(\ref{eq:localrotate1}) through
(\ref{eq:localrotate4}), it is easily derived how the $\hat{\Gamma}_i$
transform.

So far, the analysis is similar to Sec.~\ref{subsec:shakeup}. However, in the
local model, the eigenstates are local and this changes the analysis of
Sec.~\ref{subsec:shakeup} at two important points. The first one is that the
momentum shifts of $\hat{\Gamma}_{\bf q}$ do not destroy interference terms:
any final state $\ket{f}$ can be reached with any shift of ${\bf q}$. All
interference terms can in principle be present. The second point to be noted
is that, because the eigenstates are local, the only momentum dependence of
the cross section comes in through the experimental geometry, which is
reflected in the polarization factors $P_{\Gamma}$. 

When we compare the theoretical RIXS spectrum of this model to experiment, we
again have to take into account the average over the two outgoing
polarizations. Assuming again $\mathcal{M}_{\Gamma}$ are the same for all
$\Gamma$, and introducing the same broadening as before (HWHM $\gamma \approx
30$ meV phenomenological intrinsic broadening plus $27.5$ meV HWHM
experimental broadening), we obtain the spectra shown by the solid lines in
Fig.~\ref{fig:local}. 

\begin{figure}[!htp]
	\begin{center}
	\includegraphics[width=\columnwidth]{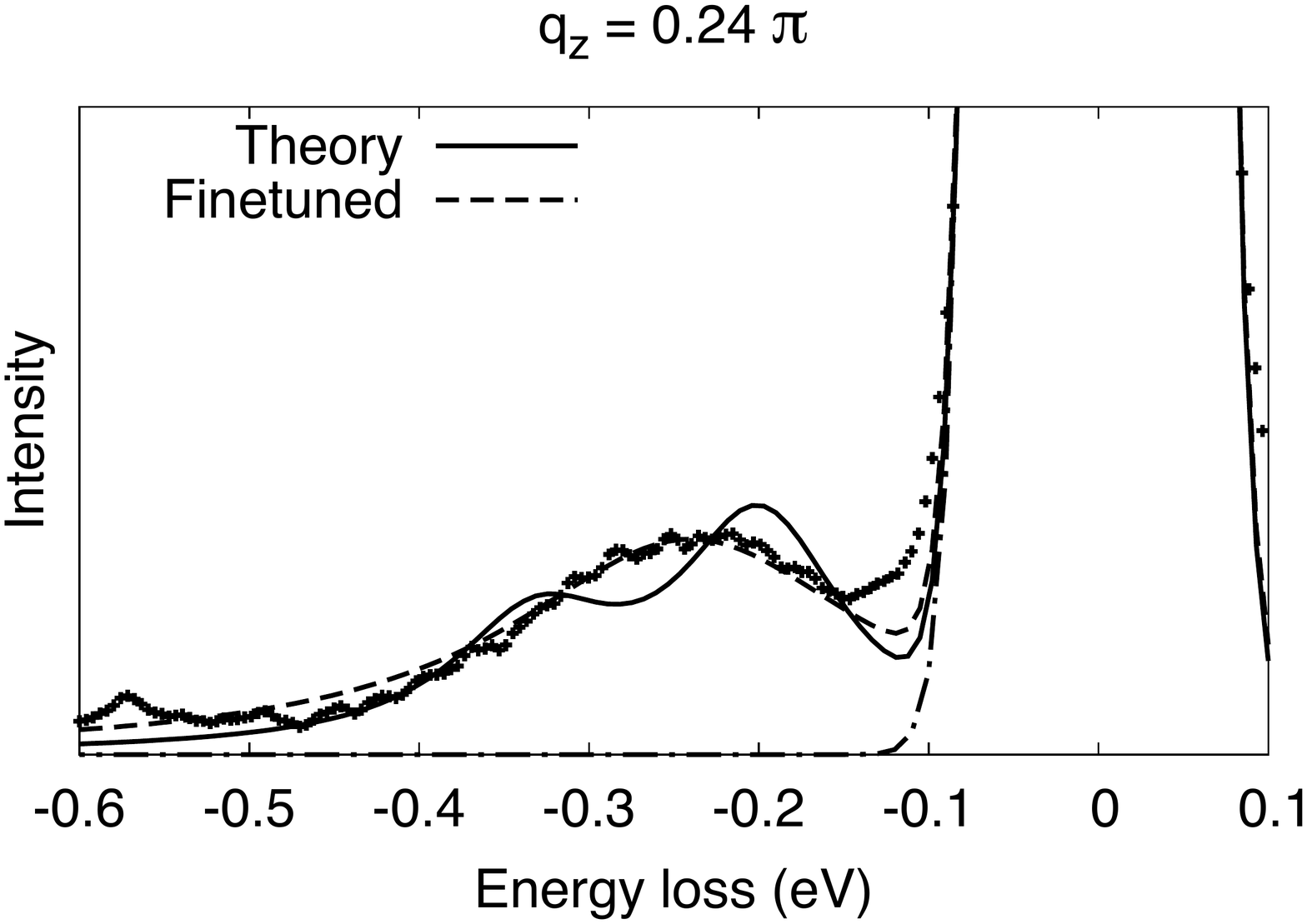}
        \includegraphics[width=\columnwidth]{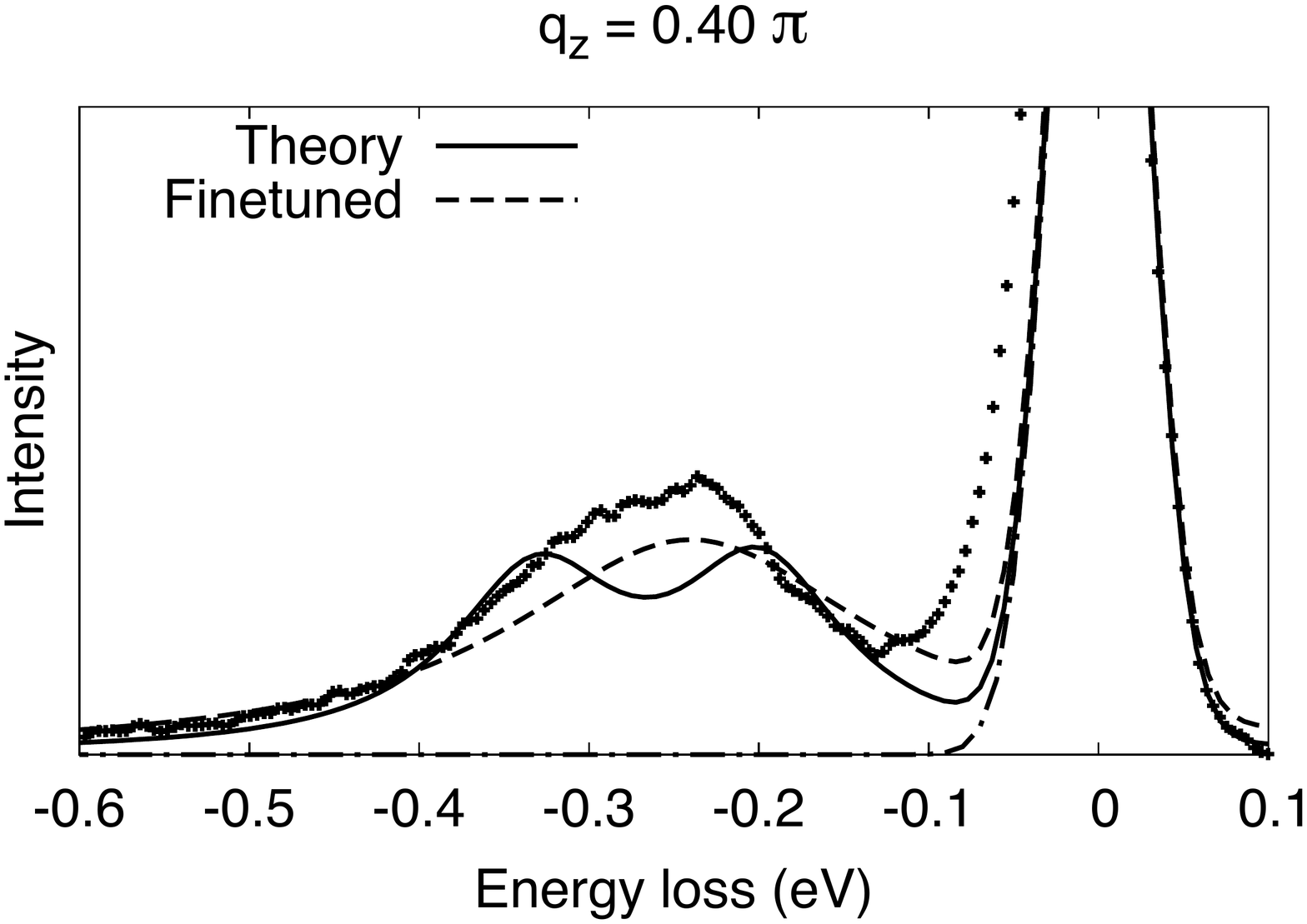}
	\includegraphics[width=\columnwidth]{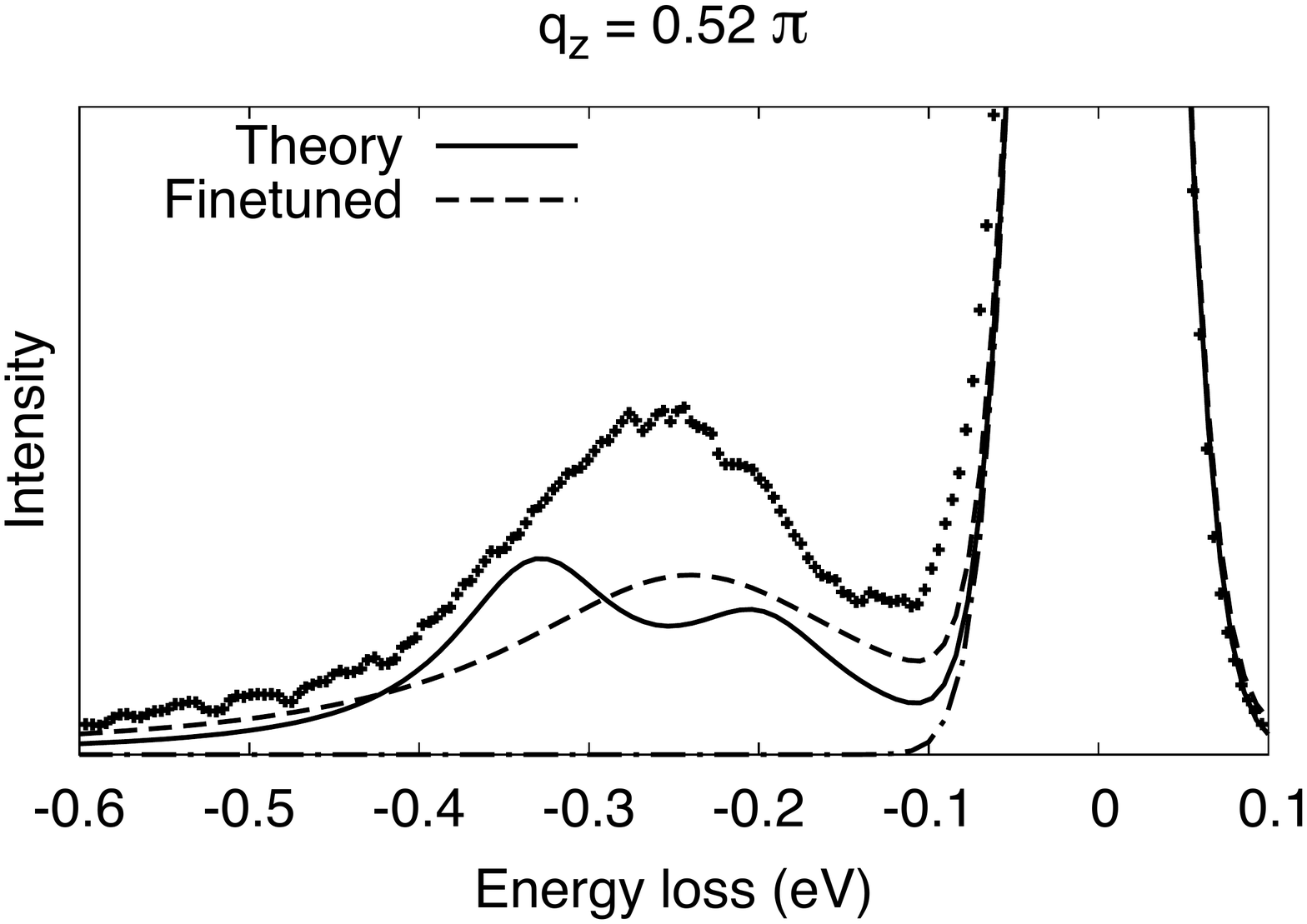}
	\caption{RIXS spectra for the local model (solid and dashed lines)
          compared to experimental data \cite{experimentalpaper}. The vector
          ${\bf q}$ is directed along the $[001]$-direction, where $q_z$ is
          indicated in the figures. The dashed curve shows the artificially
          optimized model with degenerate crystal field levels. We introduced
          a phenomenological intrinsic HWHM broadening ($\gamma \approx 30$
          meV [solid line] and $\gamma = 100$ meV [dashed line]) for the final
          states and added experimental broadening. \label{fig:local}} 
	\end{center}
\end{figure}

It is evident that the local model yields a RIXS spectrum that does
not agree well with experiment. Firstly, there is no two-peak
structure visible in the data. The presence of
a two-peak structure in the theoretical curves does not depend on the
assumption that all the $\mathcal{M}_{\Gamma}$ are equal. We may finetune the model to produce a
better fit by changing the energy levels found in
Ref.~\onlinecite{Pavarini2005} so that both crystal field transitions
have an energy of $240$ meV, and introducing a very large intrinsic
broadening of $100$ meV (see the dashed lines in
Fig.~\ref{fig:local}). But even in the artificially optimized case of
degenerate levels to produce a single peak, the intensity trend
remains in contradiction with experiment. Further, it
is impossible to tune the energy levels to optimize simultaneously the
RIXS and Raman data. Both experiments show a peak at the same energy,
while the local model theory predicts the Raman spectra (with its double
crystal field excitations) to peak at approximately
double the RIXS peak energy.

Even though we could improve the line shape by increasing $\gamma$, the
intensity gain with increasing $q_z$ cannot be reproduced in any way. In fact,
the trend is the opposite: as $q_z$ increases, the spectral weight of the
theoretical spectrum decreases (see Fig.~\ref{fig:local}). We recall that the
${\bf q}$-dependence in this case is merely due to polarization
factors Eqs.~(\ref{eq:pol_start}$-$\ref{eq:pol_end}), since in a local picture, each Ti ion
contributes independently to the cross section. This is in sharp contrast with
the superexchange picture, where the intensity has an intrinsic 
${\bf q}$-dependence because of the collective response of all the Ti ions. 

Finally, it is hard to reconcile the temperature dependence of the
experimental data with the local model. The peak is seen to broaden and lose a
large part of its spectral weight with increasing
temperature\cite{experimentalpaper}. Ascribing this broadening to phonons has
two difficulties. Jahn-Teller active phonons have energies around $30-60$ meV
and are therefore not very sensitive to temperature up to $T \approx 350-700$
K. Further, such a broadening would imply a strong orbital-lattice
coupling. This then raises the question why no structural phase transition is
seen in the titanates. We note that this is very different from, e.g.,
manganites where orbital-lattice coupling dominates. 
\end{subsection}

\begin{subsection}{Two-site processes \label{subsec:SE}}
The second term in the expansion of the effective scattering operator, Eq.~(\ref{eq:Oexpansion}),
involves two-site processes. Due to the strong multiplet effects, the
core hole potential is averaged out and becomes mainly of $A_{1g}$
symmetry. While such a potential cannot directly flip the orbitals at
the core hole site, it does affect
multi-site processes. In the case of the superexchange model, the core hole
potential effectively changes the superexchange constant $J_{SE}$
locally as discussed earlier in the context of RIXS on magnons
\cite{Brink2007,Forte2008A,Braicovich2009}. This process is illustrated in
Fig.~\ref{fig:SEmechanism}. (In principle, it is
also possible that the core hole potential modifies the orbital
interactions via the lattice vibrations.) In this section, we consider
the superexchange modulation mechanism to illustrate two-site process in
RIXS on orbital fluctuations.
\begin{figure}[!htp]
	\begin{center}
	\includegraphics[width=0.8\columnwidth]{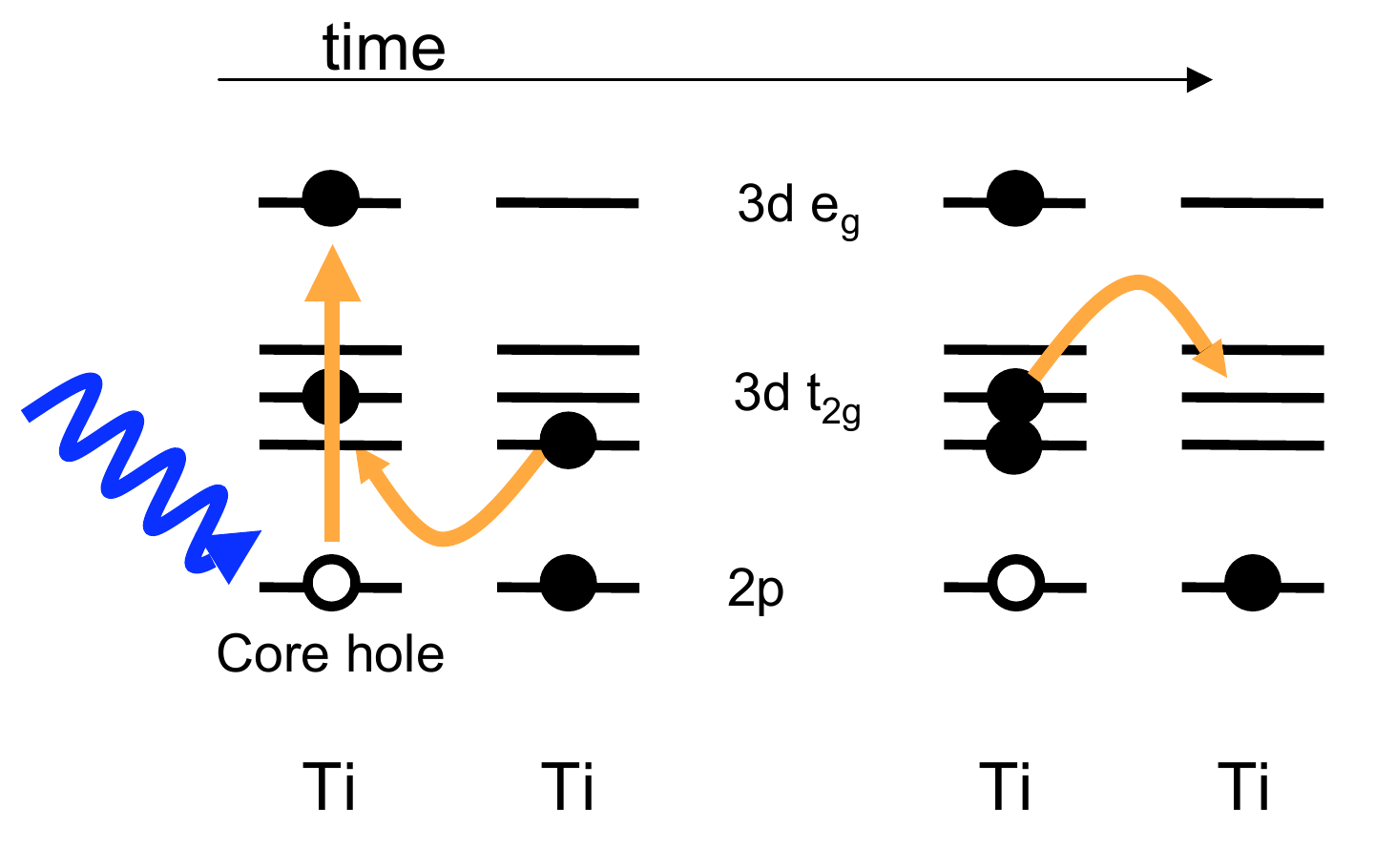}
	\caption{(Color online) In the superexchange
            model, two-site RIXS processes locally modify the
            superexchange interaction, coupling the RIXS core
          hole to the $t_{2g}$ orbitals. Shown is an orbital superexchange
          process between two neighboring Ti ions in the presence of a core
          hole. On the left, one of the ions is excited by an incoming x-ray
          photon. After that, the $t_{2g}$ electrons undergo a superexchange
          process. On the right, the virtual state of the superexchange
          process is depicted. The presence of the
            core hole frustrates the superexchange process. Instead of the usual Hubbard $U$, the energy of
          the virtual state is lowered by the presence of
          the positively charged core hole. This modifies the superexchange
          constant $J_{SE} = 4t^2/U$ at the core hole
          site. \label{fig:SEmechanism}} 
	\end{center}
\end{figure}

The superexchange modification can be derived explicitly by starting from a Hubbard model
\begin{align}
	\tilde{H} = &-t \sum_i \left( b^{\dag}_{i\pm \hat{x}} b^{\phantom{\dag}}_i + c^{\dag}_{i\pm \hat{x}} c^{\phantom{\dag}}_i + a^{\dag}_{i\pm \hat{y}} a^{\phantom{\dag}}_i + c^{\dag}_{i\pm \hat{y}} c^{\phantom{\dag}}_i + a^{\dag}_{i\pm \hat{z}} a^{\phantom{\dag}}_i \right. \nonumber \\
	&\left.+ b^{\dag}_{i\pm \hat{z}} b^{\phantom{\dag}}_i \right) + U \sum_i \left( n_{b,i} n_{c,i} + n_{a,i} n_{c,i} + n_{a,i} n_{b,i} \right) \nonumber \\
	&- U_c \sum_i p^{\phantom{\dag}}_i p^{\dag}_i \left( n_{a,i} + n_{b,i} + n_{c,i} -1\right)
\end{align}
where the last term includes the Coulomb energy $U_c$ of the core hole
attracting the $t_{2g}$ electrons. $p_i$ is the annihilation operator for 2$p$ core electrons at site $i$. We have taken the core hole potential to be of $A_{1g}$ symmetry. Doing perturbation theory to second order in $t/U_{(c)}$ ($U$ and $U_c$ are about the same order of magnitude), we obtain the superexchange Hamiltonian
\begin{equation}
	H = H_0 + \sum_{i,\delta} p^{\phantom{\dag}}_i p^{\dag}_i \left( J_2 \hat{A}^{(\gamma)}_{i,i+\delta} - J_1 \hat{n}^{(\gamma)}_{i+\delta} + {\rm const.} \right)\label{eq:Hint}
\end{equation}
with $\delta$ pointing to nearest neighbors, $n^{(c)}_i = n_{a,i} + n_{b,i}$ (the other $n^{(\gamma)}$ can be obtained by permuting the indices $a,b,c$) and
\begin{align}
	J_1 &= \frac{t^2}{U-U_c} - \frac{t^2}{U} \\
	J_2 &= \frac{t^2}{U+U_c} + \frac{t^2}{U-U_c} - \frac{2t^2}{U}
\end{align}
so that $J_1/J_2 = (1 + U/U_c)/2$. Eq.~(\ref{eq:Hint}) shows we get
the unperturbed Hamiltonian plus a contribution which is active only
if there is a core hole (in which case $p^{\phantom{\dag}}_ip^{\dag}_i
\rightarrow 1$). The $J_1$ term involves single site processes only,
and is therefore included in the general description in
Sec.~\ref{subsec:shakeup}. In the following, the $J_1$ term will be dropped.

For simplicity, polarization effects are neglected and we assume $U_c$ to be independent of the specific dipole transition. We take
\begin{equation}
	\hat{D} = \sum_i \left( e^{-i {\bf q}_{\rm in}\cdot {\bf R}_i}p^{\phantom{\dag}}_i d^{\dag}_i + e^{i {\bf q}_{\rm out}\cdot {\bf R}_i}p^{\dag}_i d^{\phantom{\dag}}_i \right) + {\rm h.c.}
\end{equation}
with $p_i$ the 2$p$ electron annihilation operator and $d_i$ the 3$d$
$e_g$ electron annihilation operator. The position of the $i^{\rm th}$
site is ${\bf R}_i$. The transfered momentum is ${\bf q} = {\bf
  q}_{\rm out} - {\bf q}_{\rm in}$. The neglected polarization
dependence could give rise to a ${\bf q}$-dependent factor in the
cross section, but will not affect the line shape for a specific ${\bf
  q}$.

The relevant energy scale for
  the excitation of orbitons in the intermediate states via
  superexchange bond modulation is $J_2$, as
  established above, as long as the core hole potential is of $A_{1g}$
  symmetry. This is the case when core level spin-orbit coupling and
  Hund's rule coupling are large compared to $J_2$: the core hole
  evolves rapidly with time and its potential's symmetry averages out
  to $A_{1g}$ before any orbitons can be excited. Because the symmetry
  is effectively cubic, bonds in all directions are affected in the
  same way. The effective scattering operator must therefore be a
  function of $\sum_\delta \hat{A}^{(\gamma)}_{i,i+\delta}$
which is of $A_{1g}$ symmetry. Non-linear
  operators like $\hat{A}^{(\gamma)}_{i,i+\delta}
  \hat{A}^{(\gamma)}_{i,i+\delta'}$
are excluded, they are expected to yield smaller contributions
because more and more distant sites are involved. In the expansion
Eq.~(\ref{eq:Oexpansion}), these come in at different orders. The only
remaining candidate for the two-site effective scattering operator is therefore 
\begin{equation}
	\hat{O}_{\bf q} = \mathcal{M}_2 \sum_{i,\delta} e^{i {\bf q}\cdot {\bf R}_i} \hat{A}^{(\gamma)}_{i,i+\delta} \label{eq:OqSE}
\end{equation}
where $\mathcal{M}_2$ is an unknown phenomenological matrix element, in the same way as 
in Sec.~\ref{subsec:shakeup}. By construction, the two-site process 
matrix element $\mathcal{M}_2$ should be proportional to $J_2$ with a constant
determined by the intermediate state dynamical susceptibilities. 
At this stage, without microscopical calculations of the single-site 
$\mathcal{M}_{\Gamma}$ (\ref{eq:Mfirstorder}) and two-site $\mathcal{M}_2$ matrix
elements, we cannot judge which coupling process dominates the observed 
RIXS on orbital excitations. Instead, we calculate two-site process 
independently and compare it with both experimental data and the results 
obtained above for single-site coupling mechanism.

As it turns out, the two-site effective scattering operator (\ref{eq:OqSE}) 
contains only two-orbiton creation terms; it does not create single orbitons 
because the orbitons are constructed in the first place to diagonalize the
Hamiltonian: all linear contributions to $\hat{A}^{(\gamma)}_{ij}$ 
in Eq.~(\ref{eq:A}) are canceled (similar to the Raman scattering
calculations above).

Using again the transformations on the orbital operators mentioned in
Ref.~[\onlinecite{Khaliullin2003}], condensing the $\psi_c$ orbital and
transforming to orbiton operators, we obtain for the two-orbiton creation part
\begin{align}
	\hat{O}^{(2)}_{\bf q} &= \mathcal{M}_2 \sum_{\bf k} \left[ f_{11} ({\bf k},{\bf q}) \alpha^{\dag}_{1,{\bf k}} \alpha^{\dag}_{1,-{\bf k}-{\bf q}} \right. \nonumber \\
	&\left. + f_{22} ({\bf k},{\bf q}) \alpha^{\dag}_{2,{\bf k}} \alpha^{\dag}_{2,-{\bf k}-{\bf q}} + f_{12} ({\bf k},{\bf q}) \alpha^{\dag}_{1,{\bf k}} \alpha^{\dag}_{2,-{\bf k}-{\bf q}} \right] \label{eq:Oq2SE}
\end{align}
where the $f_{ij}({\bf k},{\bf q})$ are lengthy functions listed in Appendix~\ref{sec:formulas}. The cross section then is
\begin{align}
	\frac{d^2\sigma^{(2)}}{d\omega d\Omega} &\propto \sum_f \abs{\bra{f}\hat{O}^{(2)}_{\bf q} \ket{0}}^2 \delta (\omega - \omega_{1/2,{\bf k}} - \omega_{1/2,{\bf k}+{\bf q}}) \nonumber \\
	&= \frac{1}{2} \sum_{\bf k} \left[ \abs{f_{11}({\bf k},{\bf q}) + f_{11}(-{\bf k}-{\bf q},{\bf q})}^2 \times \right. \nonumber \\
	&\delta (\omega - \omega_{1,{\bf k}} - \omega_{1,{\bf k}+{\bf q}}) \nonumber \\
	&+ \abs{f_{22}({\bf k},{\bf q}) + f_{22}(-{\bf k}-{\bf q},{\bf q})}^2 \times \nonumber \\
	&\delta (\omega - \omega_{2,{\bf k}} - \omega_{2,{\bf k}+{\bf q}}) \nonumber \\
	&\left. +2\abs{f_{12}({\bf k},{\bf q})}^2 \delta (\omega - \omega_{1,{\bf k}} - \omega_{2,{\bf k}+{\bf q}}) \right]. \label{eq:12susceptibility}
\end{align}
The resulting cross section for transfered momenta along the $[001]$ direction
is shown in Fig.~\ref{fig:SEspectrum}. As in the above sections, we 
introduced here by hand an energy broadening $\gamma$ of the orbiton 
states of $\gamma = 0.4\;J_{\rm orb}$.

\begin{figure}[!htp]
	\begin{center}
          \includegraphics[width=\columnwidth]{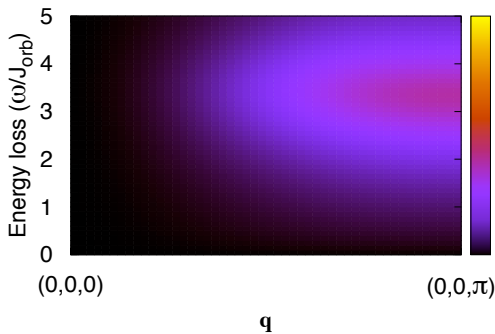}
	\caption{(Color online) RIXS spectrum for two-site processes
          within the superexchange model. The color scale denotes the
          intensity. The figure shows ${\bf q}$ running from $(0,0,0)$
          to $(0,0,\pi)$. One-orbiton creation is not allowed for the
          superexchange modulation mechanism. The intrinsic energy
          broadening $\gamma$ of the orbiton states is $\gamma =
          0.4\;J_{\rm orb} \approx 30$ meV, and the added experimental
          resolution of $27.5$ meV\cite{experimentalpaper} is
          approximately $0.34\;J_{\rm orb}$. \label{fig:SEspectrum}}
	\end{center}
\end{figure}

A few things should be noted. Firstly, the spectrum disappears at
${\bf q} = {\bf 0}$. This is clear from Eq.~(\ref{eq:OqSE}): the
scattering operator becomes proportional to the Hamiltonian Eq.~(\ref{eq:H}), giving elastic scattering only.

Secondly, the spectrum shown in Fig.~\ref{fig:SEspectrum} is calculated without taking polarization dependence into account. That could change the relative spectral weight for different ${\bf q}$'s, but does not affect the line shapes.

In Fig.~\ref{fig:SEcompare} we compare the calculated superexchange spectra for the specific ${\bf q}$ values of the experiments reported in Ref.~[\onlinecite{experimentalpaper}]. The only free parameter ($J_{\rm orb}$) gives a best fit for $J_{\rm orb} = 75$ meV. As is evident, the increase in spectral weight is qualitatively accounted for by the theory, although the theoretical curves show a much stronger increase with increasing $q_z$. We note that one factor that could diminish this discrepancy is, as stated above, the polarization factor we omitted: it could change the relative weight (but not the line shape).

Further, when we compare the theoretical line shapes with the
experimental ones, the high energy tail of the experimental data is a
bit more intense than in our calculations. This could perhaps be
accounted for if one would consider multi-orbiton
scattering. Likewise, the shoulder around $\omega = 1.5 J_{\rm orb}
\approx 110$ meV could be due to two-phonon processes. Both these
discrepancies depend on the choice of $\gamma$: a larger orbiton
damping would transfer spectral weight from the center of the
theoretical peak to its tails.

Summarizing, two-site processes can capture some of the features
seen in the RIXS data (the intensity trend with increasing momentum
transfer, and a single peak without dispersion), but the overall fit is less satisfactory
compared to the results of the single-site process shown in Fig.~\ref{fig:shakeupcompare}.

\begin{figure}[!htp]
	\begin{center}
	\includegraphics[width=\columnwidth]{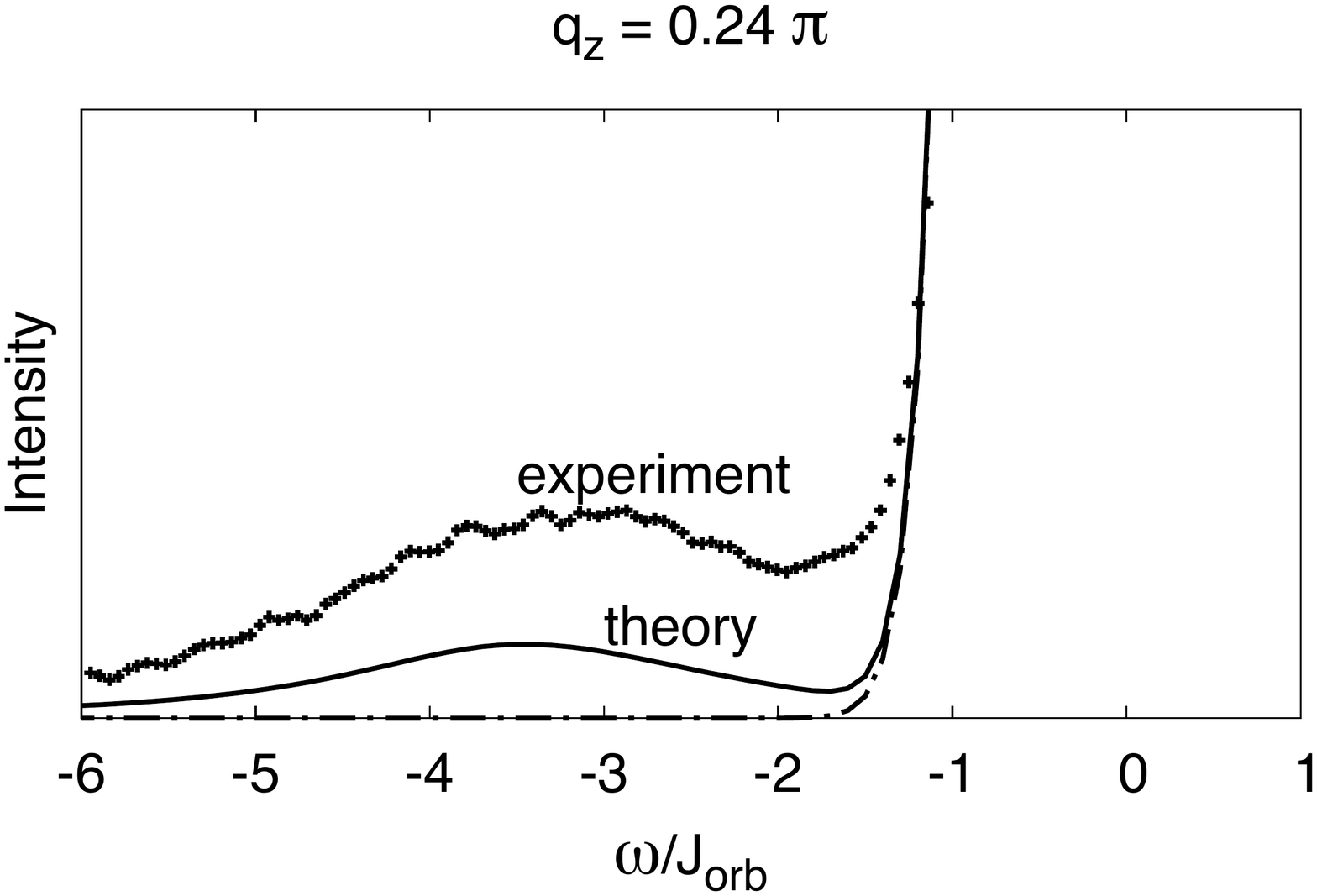}
	\includegraphics[width=\columnwidth]{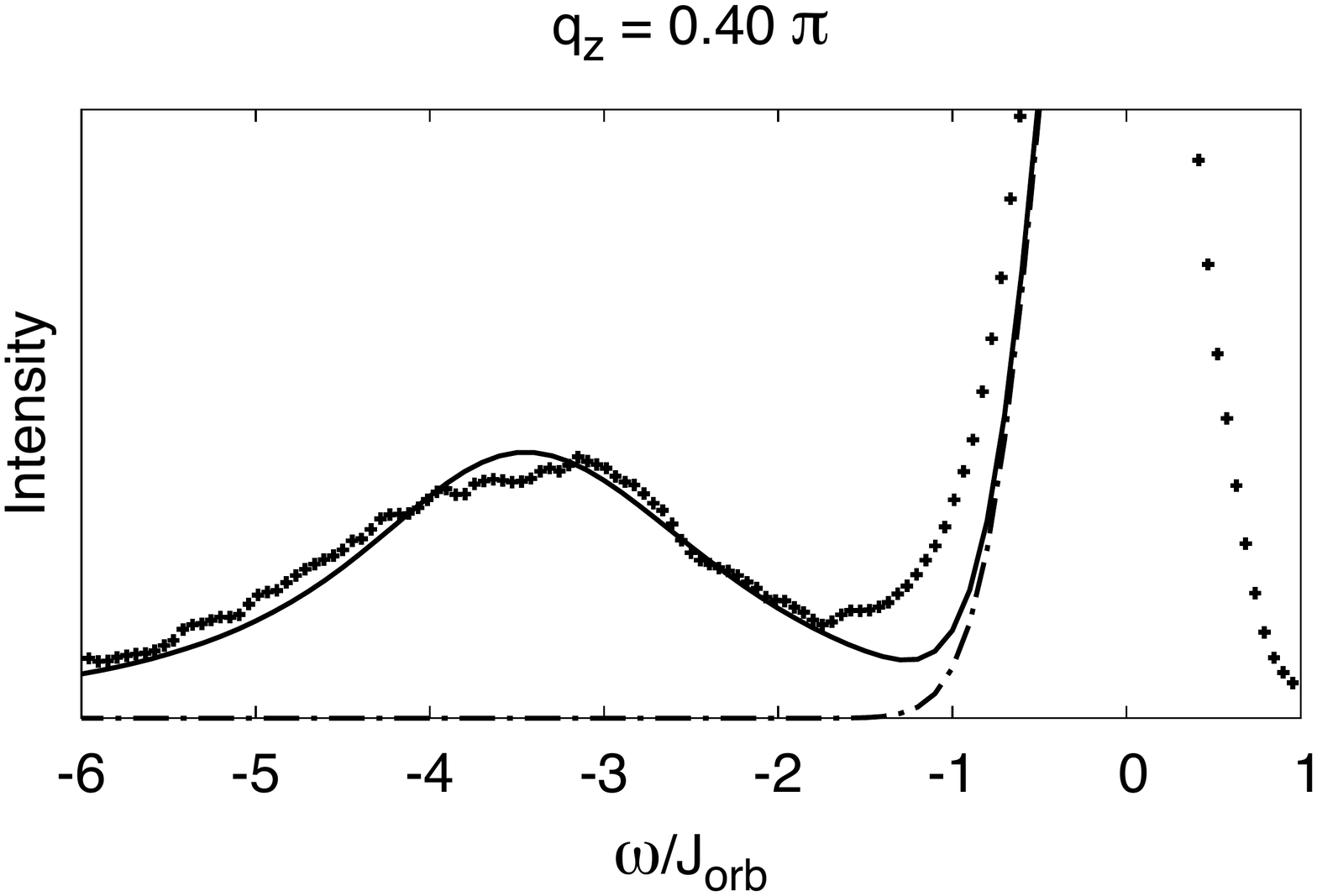}
	\includegraphics[width=\columnwidth]{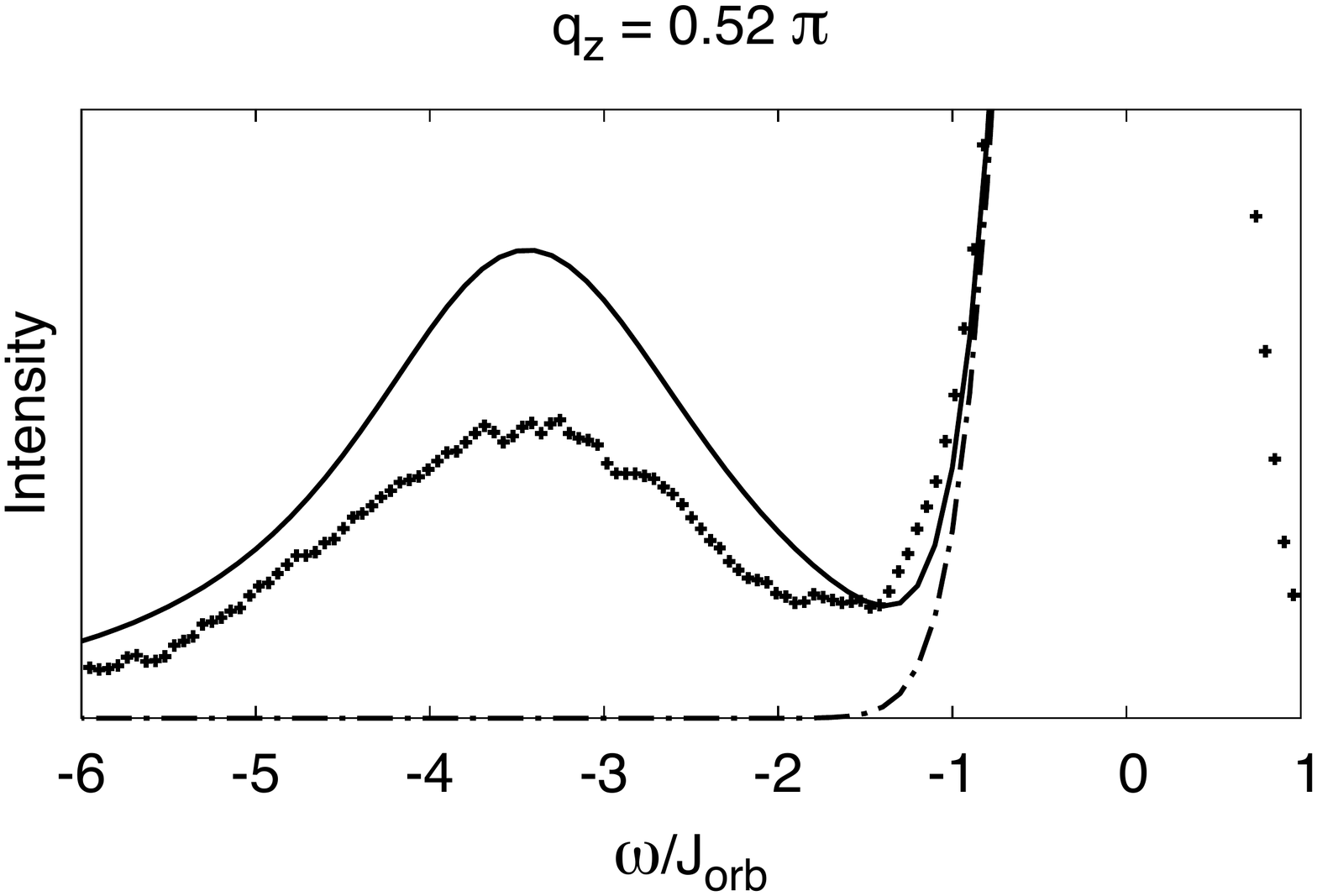}
	\caption{Theoretical RIXS spectra for two-site
            processes, calculated within the superexchange model (solid line), compared to experiment
          \cite{experimentalpaper}. ${\bf q}$ is directed along the
          $[001]$-direction, with $q_z$ as indicated. We obtain a best fit for
          $J_{\rm orb} = 75$ meV. The solid lines are cuts from the plot
          of Fig.~\ref{fig:SEspectrum}, where we added a Gaussian fit to the
          elastic peak (dash-dotted line). A phenomenological intrinsic
          orbiton broadening of $0.4 \; r_1J_{SE} \approx 30$ meV is added, as
          well as the experimental resolution of $27.5$ meV (both
          HWHM)\cite{experimentalpaper}. \label{fig:SEcompare}} 
	\end{center}
\end{figure}
\end{subsection}
\end{section}

\begin{section}{Conclusions}
We have considered two different models, widely discussed in literature to
describe orbital physics in titanites, in the context of Raman and x-ray
scattering experiments. These models correspond to two limiting cases where
the orbital ground state is dominated either by collective superexchange
interactions among orbitals or by their coupling to lattice distortions. 
The models predictions, obtained within the same level of approximations, are
compared to the experimental data on Raman (Fig.1) and on x-ray 
(Figs. 3 and 5) scattering in titanites. What is evident from this 
comparison and our detailed analysis is that the local crystal
field model of YTiO$_3$ fails to give a coherent explanation of both 
Raman and RIXS data taken together. There is no way one can get rid of the two-peak 
structure predicted for RIXS by this model without artificially
finetuning its parameters. Further, once tuned to the RIXS 
spectra, the Raman spectra will be impossible to fit with the local model
anyway, since it yields double $dd$-excitations, different from the single 
crystal field excitations in RIXS. Experimentally, however, both techniques 
show a peak at the same energy. Also, a huge anisotropy between out-of-plane 
and in-plane polarizations is predicted by the local model, which is
not observed in Raman data.
Further, the temperature dependence of the experimental data is hard to explain from a
local model: the intensity of crystal field transitions is expected to remain
unchanged. Finally, the ${\bf q}$-dependence of the RIXS-intensity is 
not reproduced by the local model; in fact, the trend is opposite. We believe
especially the last four points are robust evidence that the
$250$ meV peak seen in Raman and RIXS is not due to local $dd$-excitations. 

On the other side, the picture of collective excitations offers much better
and broad agreement with the experimental data. The general features of both
the Raman 
and RIXS data are reproduced by the superexchange model. For RIXS we presented
a phenomenological scattering operator for single and
two-site processes, evaluated within the superexchange model. Although both
single and two-site processes the general trends of the
RIXS data right, the two-site processes clearly have a too
strong ${\bf q}$-dependence of the intensity. 
The RIXS spectra obtained with the single site operator fit
the data very well, suggesting that this process of generating
orbitons might be the predominant one in the
  transition metal oxides. 
The only slight deviation from the experiments is the one-orbiton peak,
which our theory overestimates. However, we note that the theoretical
one-orbiton peak is decreased if we consider realistic conditions such as
finite temperature and residual spin-orbital quantum fluctuations, which
obstruct the orbital order and reduce $\abs{c_0}^2$, which in its turn
controls the one-orbiton spectral weight. 

Ref.~[\onlinecite{experimentalpaper}] reported that the RIXS-intensities 
in LaTiO$_3$ and YTiO$_3$ show different ${\bf q}$-dependences: While the 
intensity in YTiO$_3$ increases with ${\bf q}$, it decreases in LaTiO$_3$. 
On a qualitative level, this contrasting behavior can be understood from 
the superexchange picture as a manifestation of 
the (dynamical) Goodenough-Kanamori rules, according to which the spin 
and orbital correlations are complementary to each other. This implies 
that the spin and orbital susceptibilities are expected to behave in an
opposite fashion. Since magnetic orderings in YTiO$_3$ and LaTiO$_3$ 
are different (ferro- and antiferromagnetic, respectively), collective
response of orbitals in these coupounds are expected to be enhanced also at 
different portions of the Brillouin zone: at large ${\bf q}$ in YTiO$_3$ and, 
in contrast, at small ${\bf q}$ in LaTiO$_3$, which are complementary to the 
respective locations of their magnetic Bragg peaks. The superexchange picture 
suggests also that the ${\bf q}$-dependence of the orbiton RIXS-intensity 
should have cubic symmetry in both LaTiO$_3$ and YTiO$_3$, as follows
from their isotropic spin-wave \cite{Keimer2000,Ulrich2002} and Raman 
spectra \cite{Ulrich2006}. Future 
RIXS experiments in titanites would be useful to verify these expectations.   

A previous estimate\cite{Khaliullin2003} from neutron spin wave
data\cite{Ulrich2002} puts the orbital exchange constant $J_{\rm orb}$ at $60$
meV. In close agreement with this estimate, the theoretical Raman spectrum
fits best to experiment when $J_{\rm orb} = 65$ meV (vertex corrections may
change this number, though). Matching to a lesser degree to the estimate, 
we find for the RIXS spectra $J_{\rm orb} = 75$ and
$80$ meV for the two-site and single site processes, respectively. 

To establish the nature of the $250$ meV peak, it is of great importance to
search for the one-orbiton peak. In Raman scattering one-orbiton creation 
seems to be strongly suppressed, 
but in RIXS it would be possible to see a one-orbiton peak when
${\bf q}$ is directed maximally along the $[110]$-direction. There the
one-orbiton peak (around $\omega \approx 1.8\; J_{\rm orb} \approx 140$ meV)
is about as strong as the two-orbiton continuum, assuming
single site processes are the dominant RIXS channel, and
provided $\abs{c_0}^2$ is not too small.

To summarize, we may conclude that the existing Raman and RIXS data in
titanites are better described by the superexchange model. This implies that 
while some polarization of orbitals by static lattice distortions must be a
part of a realistic, ``ultimate'' model for titanites, the orbital
fluctuations which are intrinsic to the $t_{2g}$ orbital superexchange
process \cite{Khaliullin2000} are not yet suppressed and strong enough to 
stabilize nearly isotropic charge distributions around the Ti-ions.        

On the technical side, we believe that our semi-phenomenological 
approach to the RIXS problem which disentangles the high-energy intermediate 
state dynamics from low-energy collective excitations of orbitals and spins 
may serve as a simple and efficient tool in the theoretical description 
of Resonant Inelastic X-ray Scattering in oxides in general.
\end{section}

\begin{section}{Acknowledgements}
We would like to thank J. van den Brink for stimulating discussions that 
initiated this work. We also thank B. Keimer, C. Ulrich, and M.W. Haverkort 
for many fruitful discussions. 
L.A. thanks the Max-Planck-Institut FKF, Stuttgart, where most of the work was
done, for its hospitality. 
\end{section}

\appendix
\begin{section}{RIXS - Single site processes \label{app:shakeup}}
The angular momentum $\hat{l}$ and quadrupole operators $\hat{Q}, \hat{T}$ in Eqs.~(\ref{eq:T1u}$-$\ref{eq:T2g}) are defined as follows:
\begin{align}
	\hat{l}_x &= i(c^{\dag}b-b^{\dag}c) \label{eq:lxbare} \\
	\hat{l}_y &= i(a^{\dag}c-c^{\dag}a) \\
	\hat{l}_z &= i(b^{\dag}a-a^{\dag}b) \\
	\hat{Q}_x &= \hat{l}^2_x - \hat{l}^2_y = n_b - n_a \\
	\hat{Q}_z &= \frac{1}{\sqrt{3}} (\hat{l}^2_x + \hat{l}^2_y - 2\hat{l}^2_z) = \frac{1}{\sqrt{3}} (2n_c - n_a - n_b) \label{eq:Qzbare} \\
	\hat{T}_x &= \hat{l}_y \hat{l}_z + \hat{l}_z \hat{l}_y = -(b^{\dag}c+c^{\dag}b) \\
	\hat{T}_y &= \hat{l}_x \hat{l}_z + \hat{l}_z \hat{l}_x = -(c^{\dag}a+a^{\dag}c) \\
	\hat{T}_z &= \hat{l}_x \hat{l}_y + \hat{l}_y \hat{l}_x = -(a^{\dag}b+b^{\dag}a) \label{eq:Tzbare}
\end{align}
which are normalized by $\Tr{\hat{\Gamma}^2} = 2$. The corresponding matrices $\Gamma_{d'd}$ in Eq.~(\ref{eq:decomposeoperator}) are
\begin{eqnarray}
	\Gamma^{Q_x} = \frac{1}{2} \begin{pmatrix} -1 & 0&0 \\ 0&1&0 \\ 0&0&0 \end{pmatrix}, \;\;\; & \Gamma^{Q_z} = \frac{1}{2\sqrt{3}} \begin{pmatrix} -1 & 0&0 \\ 0&-1&0 \\ 0&0&2 \end{pmatrix},  \nonumber \\
	\Gamma^{T_x} = -\frac{1}{2} \begin{pmatrix} 0 & 0&0 \\ 0&0&1 \\ 0&1&0 \end{pmatrix}, \;\;\;	& \Gamma^{T_y} = -\frac{1}{2} \begin{pmatrix} 0 & 0&1 \\ 0&0&0 \\ 1&0&0 \end{pmatrix}, \nonumber \\
	\Gamma^{T_z} = -\frac{1}{2} \begin{pmatrix} 0 & 1&0 \\ 1&0&0 \\ 0&0&0 \end{pmatrix}, \;\;\; & \Gamma^{l_x} = \frac{1}{2} \begin{pmatrix} 0 & 0&0 \\ 0&0&i \\ 0&-i&0 \end{pmatrix}, \nonumber \\
	\Gamma^{l_y} = \frac{1}{2} \begin{pmatrix} 0 & 0&-i \\ 0&0&0 \\ i&0&0 \end{pmatrix},	\;\;\; & \Gamma^{l_z} = \frac{1}{2} \begin{pmatrix} 0 & i&0 \\ -i&0&0 \\ 0&0&0 \end{pmatrix} \label{eq:Gammadd'}
\end{eqnarray}
with the indices $d,d' = (yz,zx,xy)$ (or for polarization dependence: $\alpha,\beta = (x,y,z)$).
\end{section}

\begin{section}{Multiplet factors \label{app:multipletfactors}}
For the multiplet effect factors in Eq.~(\ref{eq:decomposepolarization}), we have
\begin{align}
	M^{A_{1g}}_{d'd} &= \sqrt{\frac{2}{3}} \left (\bra{d'} \hat{x} \ket{m} \bra{m} \hat{x} \ket{d} + \bra{d'} \hat{y}\ket{m} \bra{m} \hat{y} \ket{d} \right. \nonumber \\
	&\left.+ \bra{d'} \hat{z}\ket{m} \bra{m} \hat{z} \ket{d} \right) \label{eq:MA1g} \\
	M^{Q_x}_{d'd} &= \left( \bra{d'} \hat{y}\ket{m} \bra{m} \hat{y} \ket{d} - \bra{d'} \hat{x}\ket{m} \bra{m} \hat{x} \ket{d} \right) \\
	M^{Q_z}_{d'd} &= \frac{1}{\sqrt{3}} \left( 2 \bra{d'} \hat{z}\ket{m} \bra{m} \hat{z} \ket{d} - \bra{d'} \hat{x}\ket{m} \bra{m} \hat{x} \ket{d} \right. \nonumber \\
	&\left.- \bra{d'} \hat{y}\ket{m} \bra{m} \hat{y} \ket{d} \right) \\
	M^{T_x}_{d'd} &= -\left( \bra{d'} \hat{y}\ket{m} \bra{m} \hat{z} \ket{d} + \bra{d'} \hat{z}\ket{m} \bra{m} \hat{y} \ket{d} \right) \\
	M^{T_y}_{d'd} &= -\left( \bra{d'} \hat{z}\ket{m} \bra{m}
          \hat{x} \ket{d} + \bra{d'} \hat{x}\ket{m} \bra{m} \hat{z}
          \ket{d} \right) \\
	M^{T_z}_{d'd} &= -\left( \bra{d'} \hat{x}\ket{m} \bra{m} \hat{y} \ket{d} + \bra{d'} \hat{y}\ket{m} \bra{m}\hat{x} \ket{d} \right) \\
	M^{l_x}_{d'd} &= -i\left( \bra{d'} \hat{y}\ket{m} \bra{m} \hat{z} \ket{d} - \bra{d'} \hat{z}\ket{m} \bra{m} \hat{y} \ket{d} \right) \\
	M^{l_y}_{d'd} &= -i\left( \bra{d'} \hat{z}\ket{m} \bra{m} \hat{x} \ket{d} - \bra{d'} \hat{x}\ket{m} \bra{m} \hat{z} \ket{d} \right) \\
	M^{l_z}_{d'd} &= -i\left( \bra{d'} \hat{x}\ket{m} \bra{m} \hat{y} \ket{d} - \bra{d'} \hat{y}\ket{m} \bra{m} \hat{x} \ket{d} \right) \label{eq:Mlz}
\end{align}
Note that the position operators act on the core electrons, not the $t_{2g}$ ones. Both the core and $t_{2g}$ electrons are implied in the states $\ket{d},\ket{d'}$.
\end{section}

\begin{section}{The operators $\hat{\Gamma}$ in terms of orbitons \label{app:Oqorbitons}}
In terms of the orbiton operators, we obtain the one-orbiton creation part of $\hat{\Gamma}_{\bf q} = \sum_i e^{i{\bf q}\cdot {\bf r}_i} \hat{\Gamma}_i$ to be
\begin{widetext}
\begin{align}
	\hat{l}^{(1)}_{x,{\bf q}} &= \frac{i\abs{c_0}}{2} \sqrt{\frac{N}{3}} \left[ \left\{ (1-\sqrt{3}) u_{{\bf q}_1+{\bf q}} + (1+\sqrt{3}) v_{{\bf q}_1+{\bf q}} \right\} (\sh \theta_{1,{\bf q}_1+{\bf q}} + \ch \theta_{1,{\bf q}_1+{\bf q}} ) \alpha^{\dag}_{1,-{\bf q}_1 - {\bf q}} \right. \nonumber \\
	&+\left. \left\{ (-1-\sqrt{3}) u_{{\bf q}_1+{\bf q}} + (1-\sqrt{3}) v_{{\bf q}_1+{\bf q}} \right\} (\sh \theta_{2,{\bf q}_1+{\bf q}} + \ch \theta_{2,{\bf q}_1+{\bf q}} ) \alpha^{\dag}_{2,-{\bf q}_1 - {\bf q}} \right] \\
	\hat{l}^{(1)}_{y,{\bf q}} &= \frac{i\abs{c_0}}{2} \sqrt{\frac{N}{3}} \left[ \left\{ (1+\sqrt{3}) u_{{\bf q}_2+{\bf q}} + (1-\sqrt{3}) v_{{\bf q}_2+{\bf q}} \right\} (\sh \theta_{1,{\bf q}_2+{\bf q}} + \ch \theta_{1,{\bf q}_2+{\bf q}} ) \alpha^{\dag}_{1,-{\bf q}_2 - {\bf q}} \right. \nonumber \\
	&+\left. \left\{ (-1+\sqrt{3}) u_{{\bf q}_2+{\bf q}} + (1+\sqrt{3}) v_{{\bf q}_2+{\bf q}} \right\} (\sh \theta_{2,{\bf q}_2+{\bf q}} + \ch \theta_{2,{\bf q}_2+{\bf q}} ) \alpha^{\dag}_{2,-{\bf q}_2 - {\bf q}} \right] \\
	\hat{l}^{(1)}_{z,{\bf q}} &= -i\abs{c_0} \sqrt{\frac{N}{3}} \left[ (u_{{\bf q}_3+{\bf q}} + v_{{\bf q}_3+{\bf q}} ) (\sh \theta_{1,{\bf q}_3+{\bf q}} + \ch \theta_{1,{\bf q}_3+{\bf q}} ) \alpha^{\dag}_{1,-{\bf q}_3 - {\bf q}} \right. \nonumber \\
	&+\left. (v_{{\bf q}_3+{\bf q}} - u_{{\bf q}_3+{\bf q}}) (\sh \theta_{2,{\bf q}_3+{\bf q}} + \ch \theta_{2,{\bf q}_3+{\bf q}} ) \alpha^{\dag}_{2,-{\bf q}_3 - {\bf q}} \right] \\
	\hat{T}^{(1)}_{x,{\bf q}} &= \frac{\abs{c_0}}{6}\sqrt{N} \left[ \left\{ (1+\sqrt{3}) u_{{\bf q}_1+{\bf q}} + (-1+\sqrt{3}) v_{{\bf q}_1+{\bf q}} \right\} (\ch \theta_{1,{\bf q}_1+{\bf q}} - \sh \theta_{1,{\bf q}_1+{\bf q}} ) \alpha^{\dag}_{1,-{\bf q}_1-{\bf q}} \right. \nonumber \\
	&+\left.\left\{ (1-\sqrt{3}) u_{{\bf q}_1+{\bf q}} + (1+\sqrt{3}) v_{{\bf q}_1+{\bf q}} \right\} (\ch \theta_{2,{\bf q}_1+{\bf q}} - \sh \theta_{2,{\bf q}_1+{\bf q}} ) \alpha^{\dag}_{2,-{\bf q}_1-{\bf q}} \right] \\
	\hat{T}^{(1)}_{y,{\bf q}} &= \frac{\abs{c_0}}{6}\sqrt{N} \left[ \left\{ (1-\sqrt{3}) u_{{\bf q}_2+{\bf q}} + (-1-\sqrt{3}) v_{{\bf q}_2+{\bf q}} \right\} (\ch \theta_{1,{\bf q}_2+{\bf q}} - \sh \theta_{1,{\bf q}_2+{\bf q}} ) \alpha^{\dag}_{1,-{\bf q}_1-{\bf q}} \right. \nonumber \\
	&+\left.\left\{ (1+\sqrt{3}) u_{{\bf q}_2+{\bf q}} + (1-\sqrt{3}) v_{{\bf q}_2+{\bf q}} \right\} (\ch \theta_{2,{\bf q}_2+{\bf q}} - \sh \theta_{2,{\bf q}_2+{\bf q}} ) \alpha^{\dag}_{2,-{\bf q}_2-{\bf q}} \right] \\
	\hat{T}^{(1)}_{z,{\bf q}} &= -\frac{\abs{c_0}}{3}\sqrt{N} \left[ (u_{{\bf q}_3+{\bf q}}-v_{{\bf q}_3+{\bf q}})(\ch \theta_{1,{\bf q}_3+{\bf q}} - \sh \theta_{1,{\bf q}_3+{\bf q}} ) \alpha^{\dag}_{1,-{\bf q}_3-{\bf q}} \right. \nonumber \\
	&+\left. (u_{{\bf q}_3+{\bf q}}+v_{{\bf q}_3+{\bf q}})(\ch \theta_{2,{\bf q}_3+{\bf q}} - \sh \theta_{2,{\bf q}_3+{\bf q}} ) \alpha^{\dag}_{2,-{\bf q}_3-{\bf q}} \right] \\
	\hat{Q}^{(1)}_{x,{\bf q}} &= \abs{c_0}\sqrt{\frac{N}{3}} \left[ -(u_{\bf q}+v_{\bf q})(\ch \theta_{1,{\bf q}} - \sh \theta_{1,{\bf q}} ) \alpha^{\dag}_{1,-{\bf q}} + (u_{\bf q}-v_{\bf q})(\ch \theta_{2,{\bf q}} - \sh \theta_{2,{\bf q}} ) \alpha^{\dag}_{2,-{\bf q}} \right] \\
	\hat{Q}^{(1)}_{z,{\bf q}} &= -\abs{c_0}\sqrt{\frac{N}{3}} \left[ (u_{\bf q}-v_{\bf q})(\ch \theta_{1,{\bf q}} - \sh \theta_{1,{\bf q}} ) \alpha^{\dag}_{1,-{\bf q}} + (u_{\bf q}+v_{\bf q})(\ch \theta_{2,{\bf q}} - \sh \theta_{2,{\bf q}} ) \alpha^{\dag}_{2,-{\bf q}} \right]
\end{align}
with ${\bf q}_1 = (\pi,0,\pi),\; {\bf q}_2 = (\pi,\pi,0),\; {\bf q}_3 = (0,\pi,\pi)$. The expressions for the two-orbiton creation part of $\hat{\Gamma}_{\bf q} = \sum_i e^{i{\bf q}\cdot {\bf r}_i} \hat{\Gamma}_i$ are
\begin{align}
	\hat{l}^{(2)}_{x,{\bf q}} &= \frac{i}{\sqrt{3}} \sum_{\bf k} \left[ (vu'-uv') \ch \theta_1\; \sh \theta'_1\; \alpha^{\dag}_{1,{\bf k}} \alpha^{\dag}_{1,-{\bf k}-{\bf q}_1 - {\bf q}} + (vu'-uv') \ch \theta_2\; \sh \theta'_2\; \alpha^{\dag}_{2,{\bf k}} \alpha^{\dag}_{2,-{\bf k}-{\bf q}_1 - {\bf q}} \right. \nonumber \\
	&+\left. (uu'+vv') (\ch \theta_1 \; \sh \theta'_2 - \sh \theta_1\; \ch \theta'_2 ) \alpha^{\dag}_{1,{\bf k}} \alpha^{\dag}_{2,-{\bf k}-{\bf q}_1 - {\bf q}} \right]
\end{align}
with $u,v,\theta_1,\theta_2 = u_{\bf k},v_{\bf k},\theta_{1,{\bf k}},\theta_{2,{\bf k}}$ and $u',v',\theta'_1,\theta'_2 = u_{{\bf k}+{\bf q}_1+{\bf q}},v_{{\bf k}+{\bf q}_1+{\bf q}},\theta_{1,{\bf k}+{\bf q}_1+{\bf q}},\theta_{2,{\bf k}+{\bf q}_1+{\bf q}}$. Further, $\hat{l}^{(2)}_{y,{\bf q}}$ and $\hat{l}^{(2)}_{z,{\bf q}}$ have the same form as $\hat{l}^{(2)}_{x,{\bf q}}$ but with ${\bf q}_1$ replaced by ${\bf q}_2$ and ${\bf q}_3$ respectively. Next,
\begin{align}
	\hat{T}^{(2)}_{x,{\bf q}} &= \sum_{\bf k} \left[ \left\{ -(uu'+vv')+ \frac{1}{\sqrt{3}}(uu'-vv')+\frac{1}{3}(uv'+vu') \right\} \ch \theta_1 \; \sh \theta'_1\; \alpha^{\dag}_{1,{\bf k}} \alpha^{\dag}_{1,-{\bf k}-{\bf q}_1-{\bf q}} \right. \nonumber \\
	&+\left\{ -(uu'+vv')- \frac{1}{\sqrt{3}}(uu'-vv')-\frac{1}{3}(uv'+vu') \right\} \ch \theta_2 \; \sh \theta'_2\; \alpha^{\dag}_{2,{\bf k}} \alpha^{\dag}_{2,-{\bf k}-{\bf q}_1-{\bf q}} \nonumber \\
	&+\left.\left\{ -(uv'-vu')- \frac{1}{\sqrt{3}}(uv'+vu')-\frac{1}{3}(uu'-vv') \right\} (\ch \theta_1\; \sh \theta'_2 + \sh \theta_1\; \ch \theta'_2 ) \alpha^{\dag}_{1,{\bf k}} \alpha^{\dag}_{2,-{\bf k}-{\bf q}_1-{\bf q}} \right] \\
	\hat{T}^{(2)}_{y,{\bf q}} &= \sum_{\bf k} \left[ \left\{ -(uu'+vv')- \frac{1}{\sqrt{3}}(uu'-vv')+\frac{1}{3}(uv'+vu') \right\} \ch \theta_1 \; \sh \theta'_1\; \alpha^{\dag}_{1,{\bf k}} \alpha^{\dag}_{1,-{\bf k}-{\bf q}_2-{\bf q}} \right. \nonumber \\
	&+\left\{ -(uu'+vv')+ \frac{1}{\sqrt{3}}(uu'-vv')-\frac{1}{3}(uv'+vu') \right\} \ch \theta_2 \; \sh \theta'_2\; \alpha^{\dag}_{2,{\bf k}} \alpha^{\dag}_{2,-{\bf k}-{\bf q}_2-{\bf q}} \nonumber \\
	&+\left.\left\{ -(uv'-vu')+ \frac{1}{\sqrt{3}}(uv'+vu')-\frac{1}{3}(uu'-vv') \right\} (\ch \theta_1\; \sh \theta'_2 + \sh \theta_1\; \ch \theta'_2 ) \alpha^{\dag}_{1,{\bf k}} \alpha^{\dag}_{2,-{\bf k}-{\bf q}_2-{\bf q}} \right]
\end{align}
where in the expression for $\hat{T}^{(2)}_{y,{\bf q}}$ we replaced ${\bf q}_1$ by ${\bf q}_2$: $u',v',\theta'_1,\theta'_2 = u_{{\bf k}+{\bf q}_2+{\bf q}},v_{{\bf k}+{\bf q}_2+{\bf q}},\theta_{1,{\bf k}+{\bf q}_2+{\bf q}},\theta_{2,{\bf k}+{\bf q}_2+{\bf q}}$.
\begin{align}
	\hat{T}^{(2)}_{z,{\bf q}} &= \sum_{\bf k} \left[ \left\{ -(uu'+vv')- \frac{2}{3}(uv'+vu') \right\} \ch \theta_1 \; \sh \theta'_1\; \alpha^{\dag}_{1,{\bf k}} \alpha^{\dag}_{1,-{\bf k}-{\bf q}_3-{\bf q}} \right. \nonumber \\
	&+\left\{ -(uu'+vv')+ \frac{2}{3}(uv'+vu') \right\} \ch \theta_2 \; \sh \theta'_2\; \alpha^{\dag}_{2,{\bf k}} \alpha^{\dag}_{2,-{\bf k}-{\bf q}_3-{\bf q}} \nonumber \\
	&+\left.\left\{ -(uv'-vu')+\frac{2}{3}(uu'-vv') \right\} (\ch \theta_1\; \sh \theta'_2 + \sh \theta_1\; \ch \theta'_2 ) \alpha^{\dag}_{1,{\bf k}} \alpha^{\dag}_{2,-{\bf k}-{\bf q}_3-{\bf q}} \right]
\end{align}
where we replaced ${\bf q}_1$ by ${\bf q}_3$: $u',v',\theta'_1,\theta'_2 = u_{{\bf k}+{\bf q}_3+{\bf q}},v_{{\bf k}+{\bf q}_3+{\bf q}},\theta_{1,{\bf k}+{\bf q}_3+{\bf q}},\theta_{2,{\bf k}+{\bf q}_3+{\bf q}}$.  Finally,
\begin{align}
	\hat{Q}^{(2)}_{x,{\bf q}} &= -\frac{1}{\sqrt{3}} \sum_{\bf k} \left[ -(uu'-vv')\ch \theta_1\; \sh \theta'_1\; \alpha^{\dag}_{1,{\bf k}} \alpha^{\dag}_{1,-{\bf k}-{\bf q}} + (uu'-vv')\ch \theta_2\; \sh \theta'_2\; \alpha^{\dag}_{2,{\bf k}} \alpha^{\dag}_{2,-{\bf k}-{\bf q}} \right. \nonumber \\
	&+\left. (uv'+vu')(\ch \theta_1\; \sh \theta'_2 + \sh \theta_1\; \ch \theta'_2) \alpha^{\dag}_{1,{\bf k}} \alpha^{\dag}_{2,-{\bf k}-{\bf q}} \right] \\
	\hat{Q}^{(2)}_{z,{\bf q}} &= \frac{1}{\sqrt{3}} \sum_{\bf k} \left[ (uv'+vu')\ch \theta_1\; \sh \theta'_1\; \alpha^{\dag}_{1,{\bf k}} \alpha^{\dag}_{1,-{\bf k}-{\bf q}} - (uv'+vu')\ch \theta_2\; \sh \theta'_2\; \alpha^{\dag}_{2,{\bf k}} \alpha^{\dag}_{2,-{\bf k}-{\bf q}} \right. \nonumber \\
	&-\left. (uu'-vv')(\ch \theta_1\; \sh \theta'_2 + \sh \theta_1\; \ch \theta'_2) \alpha^{\dag}_{1,{\bf k}} \alpha^{\dag}_{2,-{\bf k}-{\bf q}} \right]
\end{align}
where in both equations we replaced ${\bf q}_1$ by ${\bf 0}$: $u',v',\theta'_1,\theta'_2 = u_{{\bf k}+{\bf q}},v_{{\bf k}+{\bf q}},\theta_{1,{\bf k}+{\bf q}},\theta_{2,{\bf k}+{\bf q}}$.
\end{widetext}
\end{section}

\begin{section}{RIXS - two-site processes with superexchange
    model\label{sec:formulas}} 

Functions $f_{11}, f_{22}$ and $f_{12}$ in
Eqs.~(\ref{eq:Oq2SE}$-$\ref{eq:12susceptibility}) are: 
\begin{widetext}
\begin{align}
	f_{11} ({\bf k},{\bf q}) &= \left[- \gamma_{3,{\bf q}}(uv'+u'v) - \gamma_{2,{\bf q}}(uu'-vv') - (1+\gamma_{1,{\bf q}})(uu'+vv')\right] (\ch \theta_1 \; \sh \theta'_1+\sh \theta_1 \; \ch \theta'_1) \nonumber \\
	&+ 2 \left[ \gamma'_1 (uu'+vv') + \gamma'_2 (uu'-vv') + \gamma'_3 (uv'+u'v) \right] (\sh \theta_1 \; \sh \theta'_1 + \ch \theta_1 \; \ch \theta'_1) \\
	f_{22} ({\bf k},{\bf q}) &= \left[\gamma_{3,{\bf q}}(uv'+u'v) + \gamma_{2,{\bf q}}(uu'-vv') - (1+\gamma_{1,{\bf q}})(uu'+vv')\right] (\ch \theta_2 \; \sh \theta'_2+\sh \theta_2 \; \ch \theta'_2) \nonumber \\
	&+ 2 \left[ \gamma'_1 (uu'+vv') - \gamma'_2 (uu'-vv') - \gamma'_3 (uv'+u'v) \right] (\sh \theta_2 \; \sh \theta'_2 + \ch \theta_2 \; \ch \theta'_2) \\
	f_{12} ({\bf k},{\bf q}) &= 2 \left[\gamma_{3,{\bf q}}(uu'-vv') + \gamma_{2,{\bf q}}(uv'+u'v)-(1+\gamma_{1,{\bf q}})(uv'-u'v)\right] (\ch \theta_1 \; \sh \theta'_2+\sh \theta_1 \; \ch \theta'_2) \nonumber \\
	&+ 4 \left[ \gamma'_1 (uv'-u'v) + \gamma'_2 (uv'+u'v) - \gamma'_3 (uu'-vv') \right] (\sh \theta_1 \; \sh \theta'_2 + \ch \theta_1 \; \ch \theta'_2)
\end{align}
where we shortened notation by writing $\theta_{1/2} = \theta_{1/2,{\bf k}},\; 	\theta'_{1/2} = \theta_{1/2,{\bf k}+{\bf q}},\; u(') = u_{{\bf k}(+{\bf q})},\; v(') = v_{{\bf k}(+{\bf q})}$ and $\gamma'_i = \gamma_{i,{\bf k}+{\bf q}}$.
\end{widetext}
\end{section}

\end{document}